\documentclass[showpacs,twocolumn,superscriptaddress,nofootinbib,dvipsnames]{revtex4-1} 

\usepackage[utf8]{inputenc}
\usepackage{hyperref}
\hypersetup{colorlinks=true,linkcolor=blue,citecolor=blue,urlcolor=blue}
\usepackage{graphicx}
\usepackage{xcolor, soul}
\sethlcolor{green}
\usepackage{dcolumn}
\usepackage{bm}
\usepackage{color}
\usepackage{enumitem}
\usepackage{mathrsfs}
\usepackage{amsmath}
\usepackage{amssymb}
\usepackage{cleveref}
\usepackage{comment}
\usepackage{orcidlink}

\usepackage{soul,xcolor}
\setstcolor{red}

\begin{document}

\title{Effect of induced frame dragging on energy extraction from axially symmetric magnetized Reissner–Nordstr\"om BH and observational benchmarks}

\author{Ikhtiyor Eshtursunov}
\email{eshtursunovikhtiyor@gmail.com}
\affiliation{New Uzbekistan University, Movarounnahr str. 1, Tashkent 100000, Uzbekistan}

\author{Chen Zhou}
\email{chenzhou@hit.edu.cn}
\affiliation{School of Physics, Harbin Institute of Technology, Harbin 150001, People’s Republic of China}

\author{Sanjar Shaymatov}
\email{sanjar@astrin.uz}
\affiliation{Institute of Fundamental and Applied Research, National Research University TIIAME, Kori Niyoziy 39, Tashkent 100000, Uzbekistan}
\affiliation{University of Tashkent for Applied Sciences, Str. Gavhar 1, Tashkent 100149, Uzbekistan}
\affiliation{National University of Uzbekistan, Tashkent 100174, Uzbekistan}

 \author{Pankaj Sheoran
 }
\email{pankaj.sheoran@vit.ac.in}
\affiliation{Department of Physics, School of Advanced Sciences, Vellore Institute of Technology, Tiruvalam Rd, Katpadi, Vellore, Tamil Nadu 632014, India}

\author{Chengxun Yuan}
\email{yuancx@hit.edu.cn}
\affiliation{School of Physics, Harbin Institute of Technology, Harbin 150001, People’s Republic of China}

\date{\today}
\begin{abstract}

In this paper, we probe the effect of induced frame dragging on the energy extraction process via magnetic reconnection in axially symmetric magnetized Reissner-Nordstr\"{o}m black hole (MRNBH) spacetime and show the influence of BH charge $Q$ and the magnetic field $B$ parameters, respectively. We start by analyzing the metric and present the distinction between the geometrical and induced ergospheres. We argue that the induced ergosphere, unlike the geometric ergosphere, is not a property of the spacetime metric but rather a particle-dependent negative energy region that changes as the charge on the particle varies. We find the region where magnetic reconnection can extract energy from the axially symmetric MRNBH and determine the conditions required for this process. We further examine the energies of the accelerated and decelerated plasma outflows and the corresponding magnetic reconnection efficiency. We show that within the parameter ranges considered, the extraction efficiency increases with the MRNBH charge $Q$ and the magnetic field parameter $B$.  For sufficiently large $B$, it reaches the efficiency comparable to that of an extremal Kerr BH case. Our results therefore show that electromagnetically induced frame dragging can support energy extraction efficiencies similar to those associated with rotation parameter. To connect with observations, we evaluate the reconnection power directly from the decelerated-plasma energy at infinity and compare it with the observed power scales of M87* and Sgr~A*. For $Q=M$, reconnection extraction appears in our scan only for $|B|M\gtrsim0.2$ ($B\sim10^{9}\,\mathrm{G}$ for M87*, $\sim10^{12}\,\mathrm{G}$ for Sgr~A*), where $P_{\mathrm{ext}}\sim10^{53}$--$10^{57}\,\mathrm{erg\,s^{-1}}$ exceeds the observed power scales by many orders of magnitude. At the weaker fields inferred from observations ($\sim10$-$10^{3}$\,G) we find no extraction, and the field-induced Blandford-Znajek (BZ) power ($\sim10^{24}$-$10^{33}\,\mathrm{erg\,s^{-1}}$ for M87*, $\sim10^{11}$-$10^{16}\,\mathrm{erg\,s^{-1}}$ for Sgr~A*) lies far below the observed jet power.

\end{abstract}
\maketitle

\section{Introduction}
\label{introduction}

The detection of gravitational waves (GWs) from binary black hole (BH) mergers \cite{Abbott16a,Abbott16b} and the reconstruction of horizon scale emission around a supermassive BH \cite{Akiyama19L1,Akiyama19L6} have opened observational windows in strong field gravity. These achievements, enabled by GW interferometers \cite{LIGOScientific:2007fwp,Accadia_2012} and the Event Horizon Telescope (EHT) \cite{Fish:2016jil}, provide a foundation for investigating BH spacetimes and their astrophysical environments. Further advances in observational capabilities \cite{KAGRA:2013rdx} offer opportunities to constrain the physical mechanisms governing these systems. These developments motivate increasingly detailed investigations of the interaction between BH geometry, electromagnetic fields, and surrounding matter/plasma. Of particular interest is the extraction of the BH rotational energy, which provides a potential energy reservoir for relativistic outflows and nonthermal particle acceleration \cite{King01ApJ,Peterson:97book,Meszaros06}.

First, the Penrose process provided a fundamental framework for extracting rotational energy from a BH \cite{Penrose:1969pc}. Within the ergoregion, a particle can split into two fragments such that one enters the horizon with negative conserved energy, allowing the other to escape with more energy than the initial particle. For the standard single-decay process around an extremal Kerr BH, the maximum efficiency is approximately $20.7\%$ under specific conditions \cite{Abdujabbarov11,Toshmatov:2014qja}. Although this mechanism establishes the extractability of BH rotational energy, its restrictive kinematic requirements motivate the consideration of electromagnetic interactions.

It must be emphasized that magnetic fields surrounding BHs can alter the motion of charged particles and facilitate access to negative-energy states. This underlies the magnetic Penrose process \cite{Bhat85,Parthasarathy86ApJ}, the Blandford–Znajek (BZ) mechanism \cite{Blandford1977}, and related electromagnetic extensions \cite{Dhurandhar1984PRD,Dhurandhar1984PRD.30.1625,Wagh85ApJ,Tursunov:2019oiq,Shaymatov24PRD.110d4042S,Xamidov24EPJC,Dadhich18MNRAS,Shaymatov24EPJC,Shaymatov22b,Khamidov25JCAP...03..053X,2026JCAP...08..035M}. Electromagnetic field effects can yield energy extraction efficiencies exceeding the limit of the purely geometric Penrose process, making magnetized rotating BHs promising candidates for efficient particle acceleration. These mechanisms have consequently been developed to gain a deeper understanding of BH energetics as a source of high-energy phenomena such as ultrahigh-energy cosmic rays \cite{PierreAuger:2017pzq,PierreAuger:2018qvk,Tursunov:2020juz}, relativistic jets \cite{Stuchlik:2015nlt,Istomin:2020oja}, and high-energy emission \cite{Ruffini:2018aiq,Kopacek:2018lgy}. Determining how the spacetime geometry and magnetic field jointly enhance energy extraction is therefore essential for assessing the astrophysical relevance of the energy extraction mechanisms.

As stated above, rotating BHs can store rotational energy that can be extracted to power high-energy astrophysical objects. The magnetic reconnection (MR) process provides one possible extraction mechanism by converting magnetic field energy into plasma motion within the ergoregion \cite{Koide08ApJ,Parfrey19PRL}. Near a rotating BH, frame dragging effect can twist magnetic field lines and help form current sheets between regions with oppositely directed magnetic fields. This is how the magnetic reconnection produces two plasma outflows: one is accelerated in the direction of BH rotation, while the other is decelerated. Under particular conditions, the decelerated outflow attains negative energy and falls into the BH, thus reducing its rotational energy. The accelerated outflow can then escape with the extracted energy. Unlike the magnetic Penrose process, which is usually described through individual charged particles, MR involves the collective motion of magnetized plasma.

However, it is worth noting that efficient extraction requires magnetic reconnection to accelerate the plasma to relativistic velocities. Slow reconnection is generally insufficient to power the most energetic sources \cite{Koide08ApJ}, whereas relativistic reconnection can rapidly convert magnetic field energy into particle motion \cite{Parfrey19PRL}. This makes MR a possible mechanism for relativistic jets and high-energy emission around BHs \cite{Blandford:1982di,Marscher:2008aa,Kotera2011ARA&A,Romero:2008zj}. However, the release energy does not necessarily imply that the energy is extracted from the BH. The energy extraction requires a negative-energy outflow to enter the horizon while its positive-energy counterpart escapes.

Later, Comisso and Asenjo \cite{Comisso21} developed a model that quantitatively evaluates this MR mechanism for rapidly rotating Kerr BH. Their approach determines the conditions for energy extraction and allows the efficiency and power to be estimated. These quantities depend on the BH spin, plasma magnetization, reconnection location, and outflow direction. The mechanism has since been investigated in various rotating BH spacetimes  \cite{Liu22ApJ,Wei22,Carleo22,Khodadi22,Wang22,Khodadi23MR_JCAP,Shaymatov24MR}. Further studies have examined how spacetime geometry affects extraction conditions and how MR compares with other energy extraction mechanisms \cite{Zhang24JCAP...07..042Z,Chen24PRD.110f3003C,Shen25PRD.111b3003S,Long25EPJC...85...26L,Rodriguez25PDU,YuChih25PRD.112j4016Y,Wang25JCAP,Zeng25PRD.112f4032Z,Zeng25PRD.112f4080Z,Cheng25EPJC...85.1130C,EshtursunovEPJC,Eshtursunov:2026-5D,EshtursunovPureLovelock}.

In this work, we investigate energy extraction through magnetic reconnection in the MRNBH spacetime \cite{Gibbons13,Shaymatov21c}. Unlike a test magnetic field, the electromagnetic field in this solution affects both the surrounding plasma and the spacetime geometry \cite{Shaymatov21c,Shaymatov:2022eyz,Shaymatov23GRG}. Its influence therefore extends beyond the motion of charged particles to the geometry governing plasma dynamics and energy extraction. Although the underlying RN BH solution is non-rotating, the interaction between its electric charge and the magnetic field parameter introduces frame dragging and can produce an ergoregion. Therefore, energy extraction can occur without introducing a spin parameter, as required for a Kerr BH surrounded by a magnetic field \cite{PhysRevD.29.2712,PhysRevD.30.1625}. From this perspective, this interaction can play a role analogous to the spin parameter of a rotating Kerr BH in inducing the frame dragging effect. Using the Comisso-Asenjo mechanism, we determine the reconnection locations and parameter ranges for which the decelerated plasma carries negative energy while the accelerated outflow carries the extracted energy outward.
 
The paper is organized as follows. In Sec.~\ref{Sec:RN}, we introduce the axisymmetric MRNBH spacetime and discuss its ergoregion and relevant properties, emphasizing their relevance to energy extraction. In Sec.~\ref{Sec:RN_MR}, we consider the application of the Comisso--Asenjo mechanism to this MRNBH solution and examine the conditions for energy extraction through magnetic reconnection. In Sec.~\ref{Sec:RN-Power-MR}, we analyze the extraction efficiency and identify the reconnection locations and the phase-space regions for which the decelerated plasma carries negative energy while the accelerated outflow gains the extracted energy. We also examine the extracted power and the overall extraction rate, highlighting their dependence on the spacetime parameters. To assess the observational relevance of our results, we estimate the reconnection power from the energy at infinity of the decelerated plasma and compare it with the observationally inferred power scales for M87* and Sgr~A* in the final part of Sec.~\ref{Sec:RN-Power-MR}. Finally, we summarize our main results in Sec.~\ref{Sec:Conclusion}. Throughout this paper, we use a system of geometric units in which we set $G_{\rm{N}}=c=1$.

\begin{figure*} 
\begin{center} \begin{tabular}{c c} 
\includegraphics[scale=0.47]{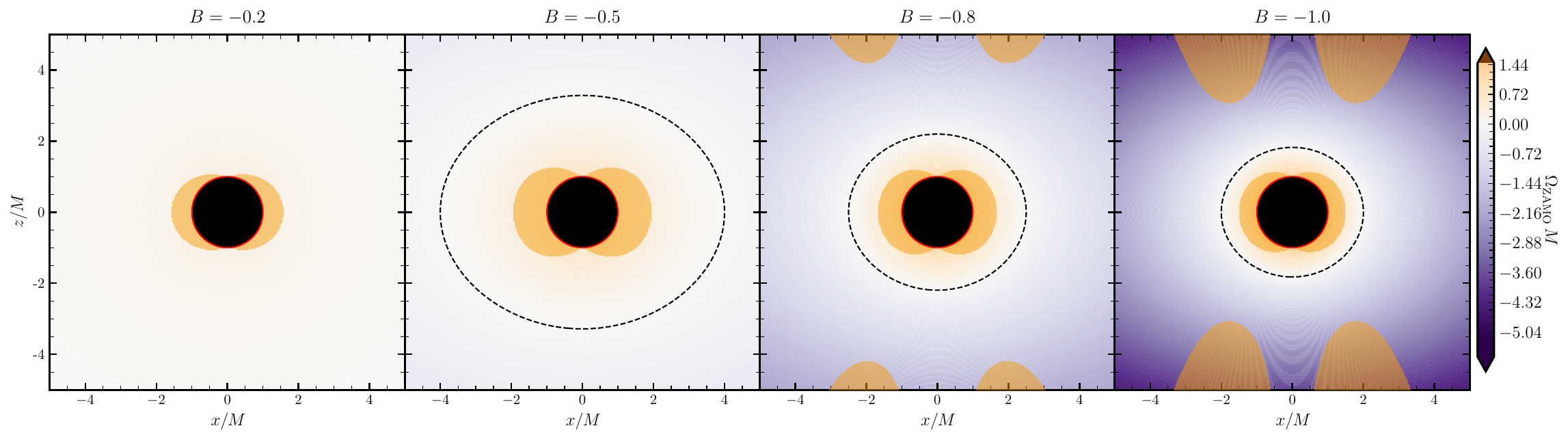}\hspace{1cm} 
\end{tabular} 
\caption{The induced frame dragging effect in a MRNBH, shown via the angular velocity $\Omega_{\mathrm{ZAMO}} = N(r,\theta)$ of zero-angular-momentum observers. The color map in the $x$-$z$ plane reveals the distribution of $\Omega_{\mathrm{ZAMO}}$ for magnetic field parameters $B = -0.2$ to $-1.0$. The yellowish shaded region corresponds to the ergoregion ($g_{tt} > 0$). The event horizon $r_+$ is shown as red curves and the black dashed curve represents $\Omega_{\mathrm{ZAMO}}=0$. Here, the parameters $M$ and $Q/M$ are set to unity.}
\label{fig:omega_zamo_ergoregion} 
\end{center}
\end{figure*}

\begin{figure}  
\includegraphics[scale=0.6]{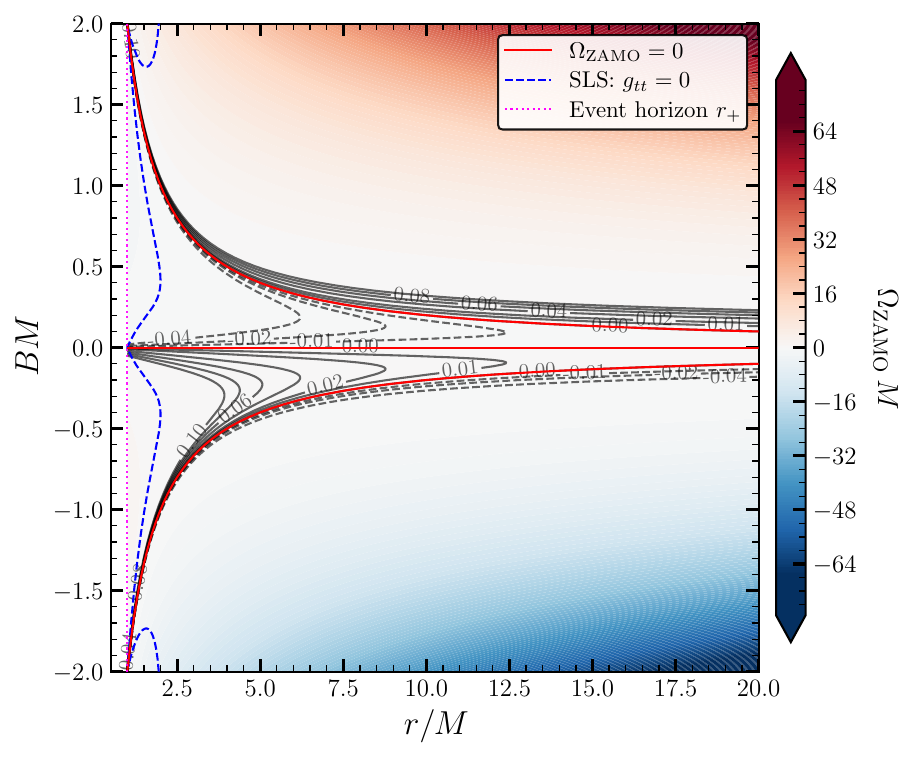}
\caption{The $\Omega_{\mathrm{ZAMO}} = N(r, \theta)$ in the $\theta = \pi/2$ plane as a function of the radial coordinate $r$ and the magnetic field parameter $B$. The color map shows the magnitude and sign of $\Omega_{\mathrm{ZAMO}}$, while the contour lines represent constant values of $\Omega_{\mathrm{ZAMO}}$. The solid and dashed contours correspond to constant positive and negative values of $\Omega_{\text{ZAMO}}$. Here, the parameters $M=1$ and $Q/M=1$ are set to unity.}
\label{fig:omega_zamo_vs_B} 
\end{figure}

\section{Axially symmetric magnetized Reissner-Nordstr\"om BH spacetime}\label{Sec:RN}

The spacetime metric describing the axially symmetric MRNBH in Schwarzschild coordinates ($t,r,\theta, \phi$) is given by \cite{Gibbons13} as
\begin{eqnarray}\label{Eq:metric} d s^2 &=& H\, \left(-F dt^2 + F^{-1}\, dr^2  + r^2 d\theta^2\right) +
      H^{-1}\, r^2\sin^2\theta\, \nonumber\\
 &&\times\left(d\phi -N dt\right)^2\, ,
  \end{eqnarray}
where  
\begin{eqnarray}\label{Eq:fhw}
F&=& 1- \frac{2M}{r} + \frac{Q^2}{r^2}\, , \\
H &=& 1 +\frac{1}{2}B^2 (r^2\sin^2\theta + 3 Q^2\cos^2\theta)
\nonumber\\&&+
  \frac{1}{16} B^4 (r^2 \sin^2\theta + Q^2\cos^2\theta)^2\, ,\\
N &=& -\frac{2Q B}{r} + \frac{1}{2} Q B^3\, r (1+F
\cos^2\theta)\, , \end{eqnarray}
with parameters $M, Q$ and $B$ corresponding to the MRNBH mass, charge and magnetic field, respectively. The above metric reduces to the RN BH one in the limit of $B\to 0$, while the Schwarzschild one in the limit of $B,Q\to 0$. It is worth noting that, interestingly, it turns out that the MRNBH also causes axially symmetric spacetime due to the existence of a magnetic field without any rotation. It is a very interesting property of MRNBH.  

Interestingly, the MRNBH's horizon has the form as 
\begin{eqnarray}
r_{\pm}=M\pm\sqrt{{M}^2-Q^{2}}\, ,
\end{eqnarray}
which is the same as the one for RN BH. Apparently, the magnetic field parameter has no effect on the BH horizon.

Due to stationary and axial symmetry, there are two Killing vectors in the magnetized BH spacetime considered here and they are defined by 
$$\xi^{\mu}_{(t)}=\left(\frac{\partial}{\partial t}\right)^{\mu}\, \mbox{~~~and~~~}\\
\xi^{\mu}_{(\phi)}=\left(\frac{\partial}{\partial \phi}\right)^{\mu}\, ,
$$
The above two killing vectors, respectively, refer to a stationary and an axisymmetry of the spacetime, and thus there exist two conserved quantities, i.e. the energy and angular momentum for a particle along with its mass $m$. 

An important feature of the MRNBH is the appearance of a non-vanishing $g_{t\phi}$ component even though the underlying BH is non-rotating. In fact, from Eq.~(\ref{Eq:metric}), the $t-\phi$ sector can be written as

\begin{equation}
H^{-1}r^2\sin^2\theta\left(d\phi-Ndt\right)^2,
\end{equation}
from which one obtains
\begin{equation}
g_{t\phi}=-H^{-1}r^2\sin^2\theta\,N,
\qquad
g_{\phi\phi}=H^{-1}r^2\sin^2\theta.
\end{equation}
Consequently, the angular velocity of zero-angular-momentum observers (ZAMOs) is given by
\begin{equation}
\Omega_{\rm ZAMO}
=-\frac{g_{t\phi}}{g_{\phi\phi}}
=N(r,\theta).
\end{equation}
Thus, although the BH has no intrinsic angular momentum, the magnetic field parameter together with the electric charge induces a non-trivial angular dragging of inertial frames.

Using Eq.~(\ref{Eq:fhw}), this angular velocity is explicitly given by
\begin{equation}
\Omega_{\rm ZAMO}
=
-\frac{2QB}{r}
+\frac{1}{2}QB^3r
\left(1+F\cos^2\theta\right).
\end{equation}
Curiously enough, the two terms in $\Omega_{\rm ZAMO}$ have different radial dependences. Specifically, the first term, which scales like $QB/r$, is dominant near the BH. Whereas the second term, which is $\frac{1}{2}QB^3 r (1+\cos^2\theta)$ in the asymptotic limit of $r$ ($F\rightarrow1$), becomes more significant at larger radial distances and represents the effect of the magnetic-field background. 
Therefore, it is worth pointing out that the induced angular dragging need not have the same direction throughout the spacetime. Also, note that the exact behavior of the second term is modified by the $r$-dependence of $F$, which gives the additional subleading contributions $\propto 1/r$ and $\propto 1/r^2$ that are relevant at intermediate radii.

In particular, for the case $QB\neq0$, a surface may exist on which $\Omega_{\rm ZAMO}=0$. This surface is determined by
\begin{equation}
B^2r^2\left(1+F\cos^2\theta\right)=4.
\end{equation}
Across this surface, the sign of $\Omega_{\rm ZAMO}$ changes, indicating a reversal in the direction of the induced angular dragging. This novel feature of the MRNBH, in which the magnetic field modifies not only the radial structure of the spacetime but also its local rotational properties despite the absence of intrinsic BH spin makes it interesting compared to its counterpart (i.e., standard RN BH).

Let us now define the static-limit surface (SLS) for which the Killing
vector $\xi^{\mu}_{(t)} = \partial/\partial t$ becomes null. From the
metric in Eq.~(\ref{Eq:metric}), the $tt$ component is
\begin{equation}
g_{tt}
=
-HF+H^{-1}r^2\sin^2\theta\,N^2,
\end{equation}
and the static-limit surface is consequently determined by
\begin{equation}
g_{tt}=0,
\end{equation}
or in explicit form reads as 
\begin{eqnarray}
N^2r^2\sin^2\theta-H^2F=0.
\end{eqnarray}
The existence and properties of SLSs and ergoregions in
magnetized spacetimes have been investigated in~\cite{Gibbons13}. In particular, the ergoregion and its implications
for the Penrose process in the MRNBH
spacetime were studied in~\cite{Shaymatov:2022eyz}, while the electrical
Penrose process in the same spacetime has recently been revisited in~\cite{Baez:2026ich}.

For the geometrical Penrose process, the relevant ergoregion is the
region in which the stationary Killing vector becomes spacelike, namely
\begin{equation}
g_{tt}>0.
\end{equation}
Thus, the geometrical ergoregion is determined entirely by the
background MRNBH geometry and is bounded by
the corresponding SLS. In the present spacetime, the
ergoregion need not be connected to the BH horizon and may
contain disconnected, remote regions, as discussed in~\cite{Shaymatov:2022eyz}. It is therefore useful to distinguish this
geometrical ergoregion from the negative-energy regions that may arise
for charged particles through their electromagnetic interaction with
the background field.

In Fig.~\ref{fig:omega_zamo_ergoregion}, we show the geometrical
ergoregion for the extremal axially symmetric MRNBH on the $x$-$z$ plane for various values
of the magnetic-field parameter $B$. As can be seen from this figure and the relevant figures in \cite{Shaymatov:2022eyz},
the ergoregion can extend to large distances from the BH in
both the positive and negative $z$ directions. For sufficiently small
$|B|$, the ergoregion may remain disconnected from the BH
horizon, whereas for sufficiently large $|B|$ it merges with the
BH horizon. This behavior is consistent with the ergoregion
structure previously reported for MRNBHs \cite{Shaymatov:2022eyz} in detail. The corresponding behavior of
$\Omega_{\rm ZAMO}$ is shown in
Fig.~\ref{fig:omega_zamo_vs_B}. As discussed in our earlier work \cite{Shaymatov:2022eyz}, the
size of the ergoregion exhibits a non-monotonic dependence on the
magnetic-field parameter: it initially increases with $|B|$, reaches an
optimal value, and subsequently decreases for larger $|B|$.
Interestingly, a qualitatively similar non-monotonic behavior is also
observed in $\Omega_{\rm ZAMO}$, as illustrated in
Fig.~\ref{fig:omega_zamo_vs_B}. This feature highlights the
non-trivial effect of the magnetic field on the local
rotational properties of the spacetime, despite the absence of
intrinsic angular momentum of the seed RN BH.

For the conventional Penrose process, the existence of a geometrical
ergoregion, characterized by $g_{tt}>0$, is essential because it permits
negative values of the conserved energy associated with the stationary
Killing vector. In the present analysis, we focus on the extremal
MRNBH, corresponding to
$Q=M$ (and, in the dimensionless units adopted in the figures,
$Q=M=1$). This choice is motivated by the fact that the efficiency of
the conventional Penrose process generally reaches its maximum in the
extremal limit, where the Cauchy and event horizons coincide
\cite{1985JApA....6...85B}. It therefore provides a useful limiting case
for investigating the maximum impact of the magnetic field on the
geometrical ergoregion and the associated energy-extraction process.

\medskip
\noindent\textit{Geometrical versus induced ergosphere:} It is
important to distinguish the geometrical ergoregion discussed above
from the negative-energy region relevant to the electrical Penrose
process for charged particles. The geometrical ergoregion is defined by
the causal character of the stationary Killing vector,
\begin{equation}
\mathcal{E}_{\rm geo}
=
\left\{
x^\mu\,\big|\,g_{tt}>0
\right\},
\end{equation}
and is therefore a property of the background spacetime. Equivalently,
it corresponds to the region in which
\begin{equation}
\xi^\mu_{(t)}\xi_{(t)\mu}>0.
\end{equation}
Its boundary is the static-limit surface, where
\begin{equation}
g_{tt}=0.
\end{equation}
In contrast, for a charged particle the conserved energy associated
with the stationary Killing field contains an electromagnetic
contribution and can be written as
\begin{equation}
E=-p_t-qA_t,
\end{equation}
where $q$ is the particle charge and $A_\mu$ is the electromagnetic
four-potential. Consequently, the condition for a charged particle to
possess negative canonical energy need not coincide with the
geometrical condition $g_{tt}>0$.

This distinction has recently been emphasized in the analysis of the
electrical Penrose process in the MRNBH
spacetime by B\'aez, Bret\'on, and Cabrera-Munguia \cite{Baez:2026ich},
which revisits and extends the energy-extraction analysis of~\cite{Shaymatov:2022eyz}. In particular, for a static charged BH, the electromagnetic interaction can produce negative-energy
states even in the absence of a geometrical ergoregion. In their
analysis of the RNdS BH spacetime, the
boundary of such a negative-energy region is obtained from the
effective-potential condition
\begin{equation}\label{eq:N_E}
E=V_+=0.
\end{equation}
For vanishing angular momentum, $L=0$, this condition takes the form
\begin{equation}\label{eq:in_ergo}
g_{tt}+(eA_t)^2=0,
\qquad
e\equiv\frac{q}{m}.
\end{equation}
Unlike the geometrical static-limit condition $g_{tt}=0$, this
condition explicitly depends on the charge-to-mass ratio of the
probing particle. Hence, the corresponding surface is not an
ergosurface in the strict geometrical sense, since it is not determined
solely by the spacetime metric. Rather, it identifies the boundary of
a particle-dependent negative-energy region.

We therefore refer to this latter region as an \emph{induced
ergosphere} to clearly distinguish it from the geometrical ergosphere.
More precisely, we define
\begin{equation}
\mathcal{E}_{\rm geo}
=
\left\{
x^\mu\,\big|\,g_{tt}>0
\right\}
\end{equation}
as the geometrical ergoregion, whereas the induced ergosphere
associated with a charged probe is determined by the corresponding
negative-energy condition Eq. (\ref{eq:N_E}). For $L=0$, its boundary is consequently given by the Eq. (\ref{eq:in_ergo}). As the latter is explicitly probe-dependent through $e=q/m$. Thus,
particles with different charge-to-mass ratios can have different
boundaries of their negative-energy regions in the same background
spacetime. In this sense, we would like to emphasise that the induced ergosphere should not be
interpreted as an additional geometrical ergoregion of the spacetime.

For the MRNBH geometry studied here, the
region shown in Fig.~\ref{fig:omega_zamo_ergoregion} is instead a
genuine geometrical ergoregion because it is obtained directly from
$g_{tt}>0$. Its existence is therefore a property of the magnetized
background and is independent of the charge-to-mass ratio of a
particular probing particle. The magnetic field modifies the
spacetime itself and generates a non-vanishing $g_{t\phi}$, resulting
in the induced angular dragging characterized by
$\Omega_{\rm ZAMO}=N(r,\theta)$. Consequently, although the efficiency
of energy extraction by charged particles depends on their
electromagnetic coupling, the geometrical ergoregion itself does not.

This distinction is particularly relevant when comparing the
conventional Penrose process with the electrical Penrose process. The
geometrical condition $g_{tt}>0$ characterizes the region in which the
stationary Killing vector becomes spacelike and provides the
background ergoregion, whereas the electromagnetic contribution to
the particle's conserved energy determines whether a particular
charged fragment can occupy a negative-energy state. The two notions
should therefore be kept conceptually separate: the former is a
\emph{geometrical ergosphere}, while the latter represents a
\emph{probe-dependent induced ergosphere}. This distinction is
important for a consistent interpretation of energy extraction by
neutral and charged particles in MRNBH spacetimes.

The simultaneous investigation of $\Omega_{\rm ZAMO}$ and $g_{tt}$ therefore provides a direct characterization of the induced rotational structure and the ergoregion of the magnetized BH. In particular, the $\Omega_{\rm ZAMO}=0$ surface identifies the region in which the direction of magnetic-field-induced frame dragging changes, while $g_{tt}=0$ determines the boundary beyond which static observers cannot exist.

The electromagnetic field around the magnetized BH can be defined by 
\begin{eqnarray}\label{Eq:4-vec}
A &=& A_t dt + A_\phi (d\phi-\omega dt)\, , 
\end{eqnarray}
with the nonvanishing vector potential components (see details \cite{Shaymatov21c})  
 \begin{eqnarray}
A_t &=& -\frac{Q}{r} +\frac{3}{4} Q B^2 r\, (1+ F\cos^2\theta)\, ,\nonumber\\
A_\phi &=& \frac{2}{B} - H^{-1}\Big[\frac{2}{B} +
\frac{1}{2} B(r^2\sin^2\theta + 3 Q^2
\cos^2\theta)\Big]\, . \nonumber\\
\end{eqnarray}

The relevance of $\Omega_{\rm ZAMO}$ in the context of magnetic reconnection and energy extraction becomes evident when we consider the electric field induced in the plasma frame. In a magnetized plasma surrounding the BH, the presence of a non-zero $\Omega_{\rm ZAMO}$ leads to a motional electric field given by
\begin{equation}
\mathbf{E} = -(\mathbf{v} \times \mathbf{B}) = -(\boldsymbol{\Omega}_{\rm ZAMO} \times \mathbf{r}) \times \mathbf{B},
\end{equation}
which in the poloidal plane can be expressed as
\begin{equation}
E_r = \Omega_{\rm ZAMO} \, r \sin\theta \, B_\theta, \qquad E_\theta = -\Omega_{\rm ZAMO} \, r \sin\theta \, B_r.
\end{equation}
This induced electric field can become comparable to the gravitational field near the BH, particularly in the ergoregion where $\Omega_{\rm ZAMO}$ is large. The presence of such electric fields is crucial for driving magnetic reconnection, as they can accelerate particles and create current sheets where magnetic field lines of opposite polarity approach each other.

More quantitatively, the reconnection rate in such a system is determined by the dimensionless parameter
\begin{equation}
\sigma_{\rm rec} \sim \frac{E}{B} \sim \Omega_{\rm ZAMO} \, r \sin\theta,
\end{equation}
which represents the ratio of the induced electric field to the magnetic field strength. This parameter can become of order unity in the ergoregion, indicating that reconnection proceeds at the fast rate $\sim 0.1\, v_A$, where $v_A$ is the Alfv\'en speed. The figure of $\Omega_{\rm ZAMO}$ distribution (Fig.~\ref{fig:omega_zamo_ergoregion}) therefore directly shows where the induced electric field is strongest and, consequently, where magnetic reconnection is most efficient.

Furthermore, the surface $\Omega_{\rm ZAMO}=0$ is particularly interesting for reconnection because it separates regions of opposite frame-dragging direction. At this surface, the magnetic field lines experience a strong shear, creating a natural site for current sheet formation. The energy released during reconnection at this surface can be estimated from the difference in the electromagnetic energy across the surface, which is proportional to
\begin{equation}
\Delta \mathcal{E} \sim \frac{B^2}{8\pi} \, \Delta V \, \left(\frac{\Delta \Omega}{\Omega}\right)^2,
\end{equation}
where $\Delta V$ is the volume of the reconnection region and $\Delta \Omega$ is the change in $\Omega_{\rm ZAMO}$ across the surface. This shear-induced reconnection provides an efficient mechanism for extracting the BH's electromagnetic energy.

This induced rotational structure is particularly relevant for the energy-extraction processes considered below, since the existence and spatial distribution of the ergoregion determine the conditions under which negative-energy plasma states can occur and, consequently, whether magnetic reconnection can efficiently extract energy from the system. In Figs.~\ref{fig:omega_zamo_ergoregion} and \ref{fig:omega_zamo_vs_B}, we plot $\Omega_{\rm ZAMO}$ to visualize the regions where frame dragging is significant and to identify the $\Omega_{\rm ZAMO}=0$ surfaces. Fig.~\ref{fig:omega_zamo_ergoregion} shows the spatial distribution of $\Omega_{\rm ZAMO}$ together with the ergoregion, demonstrating that the ergoregion always lies in regions of strong frame dragging, while Fig.~\ref{fig:omega_zamo_vs_B} shows how $\Omega_{\rm ZAMO}$ varies with the magnetic field parameter $B$, revealing the $\Omega_{\rm ZAMO}=0$ surfaces where the direction of frame dragging reverses. These plots are essential for understanding where magnetic reconnection can occur and how the reconnection rate depends on the magnetic field strength.

\section{Application of magnetic reconnection mechanism}\label{Sec:RN_MR}

\begin{figure} 
\centering
\includegraphics[scale=0.4]{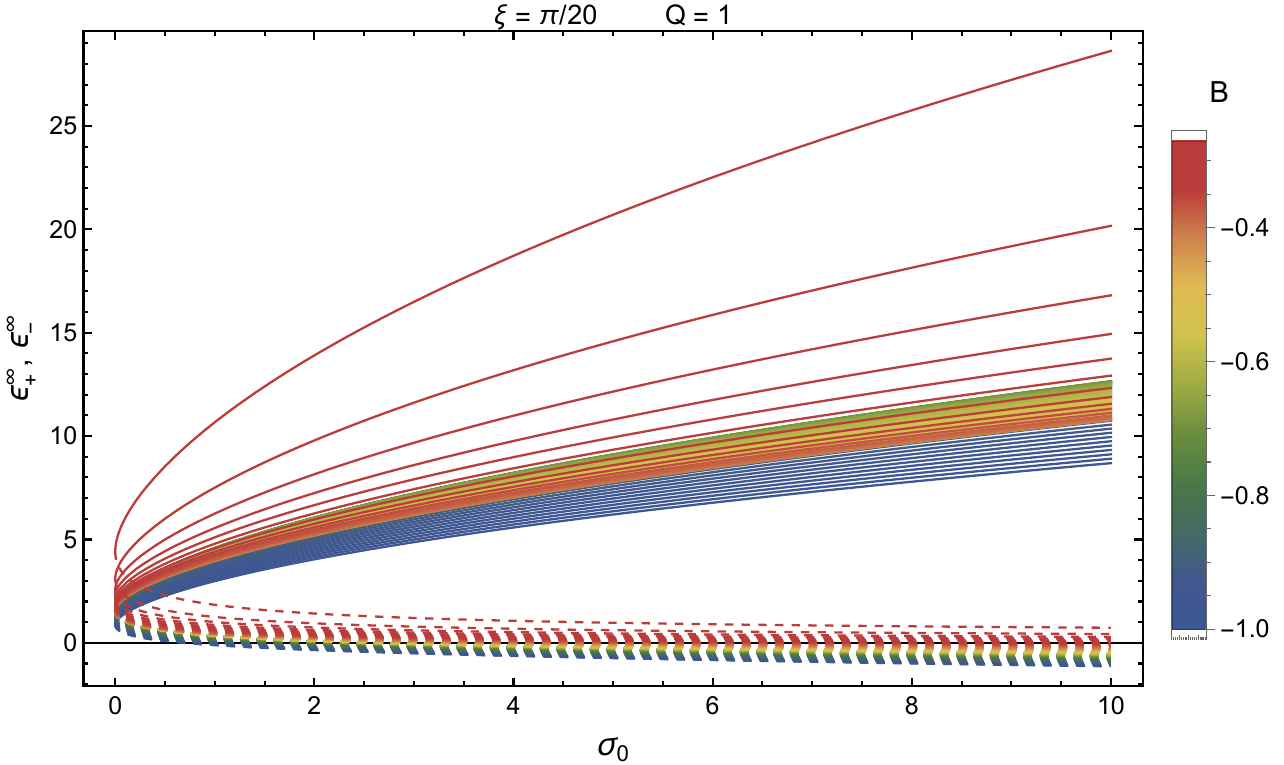}\hspace{1cm} 
\caption{\label{Fig:energyatinfsigma} Energies at infinity of the accelerated and decelerated plasma, \(\epsilon_+^\infty\) and \(\epsilon_-^\infty\), as functions of the magnetization parameter \(\sigma_0\) for axially symmetric MRNBH. Solid and dashed curves represent \(\epsilon_+^\infty\) and \(\epsilon_-^\infty\), respectively, while the color bar indicates the magnetic field parameter \(B\). The magnetic field parameter \(B\) is varied from \(-1\) to \(0\) in steps of \(0.01\). Only values satisfying the conditions for timelike circular motion are shown. Here, M is set to unity.}
\end{figure}
\begin{figure*}
\resizebox{\textwidth}{!}{
\begin{tabular}{cc}
  \includegraphics[scale=0.45]{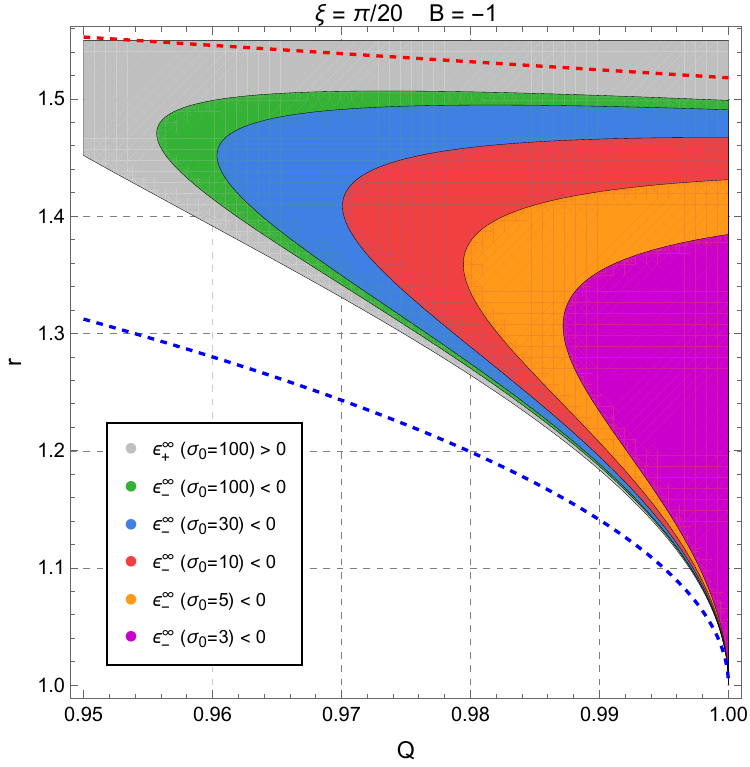}\hspace{-0.1cm}
  \includegraphics[scale=0.45]{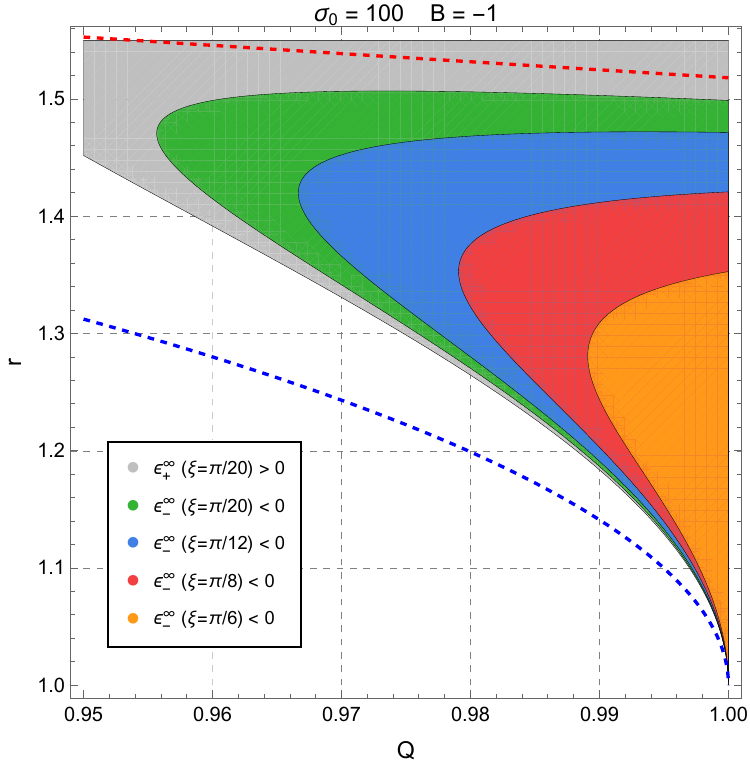}\hspace{-0.1cm}
  \end{tabular}
  }
 \caption{\label{fig:region1} For axisymmetric MRNBH the allowed energy-extraction regions for accelerated($\epsilon_+^{\infty}>0$) and decelerated($\epsilon_-^{\infty}<0$) plasma states are provided in the $Q-r$ plane. In the left panel, the regions are represented for fixed orientation angle value $\xi=\pi/20$ and varying quantities of plasma magnetization parameter(3-100). In the right panel, the results are illustrated for fixed plasma magnetization $\sigma_0=100$ and differing values of the orientation angle values($\pi/20-\pi/6$). Red and blue dashed lines represent the boundaries of ergoregions and outer event horizons, respectively. In both results B magnetic field parameter is set to -1.}
\end{figure*}

Magnetic reconnection mechanism~\cite{Comisso21} can be magnificent way to extract BHs energy by converting them into plasma kinetic energy and accelerating them in extreme astrophysical environments. Applying this mechanism to different BH geometries can help to understand the effects of BH spacetimes to the efficiency of energy extraction processes. Specifically, axisymmetric MRNBH can be interesting environment to test energy extraction through magnetic reconnection, because, in this case, the surrounding spacetime and plasma dynamics can be influenced by both the BH charge and the magnetic field. Therefore, extending the magnetic reconnection mechanism in these environments may give thorough explanation of how the BH charge and magnetic field influence the required conditions of efficient energy extraction. We analyze plasma energetics extracted from the BH in locally flat ZAMO frame (zero angular momentum observer). The plasma in our study is considered bulk plasma. Here, the bulk plasma describes the collective flow of the magnetized plasma surrounding the BH. The magnetic reconnection occurs in the equatorial plane ($\theta=\pi/2$) of the BH. We introduce the ZAMO frame and transformations of the Boyer-Lindquist coordinates to the ZAMO as:
  \begin{eqnarray}
      ds^2=-d\hat{t}^2+\sum_{i=1}^3(d\hat{x^i})^2=\eta_{\mu \nu}\hat{dx^{\mu}}\hat{dx^{\nu}}\, . 
  \end{eqnarray}
\begin{eqnarray}
d\hat{t}=\alpha dt \mbox{~~and~~} d\hat{x^i}=\sqrt{g_{ii}}dx^i-\alpha \beta^{i}dt\, ,
\end{eqnarray}
here $\alpha$ is the lapse function $\alpha=\sqrt{-g_{tt}+\frac{g_{\phi t}^2}{g_{\phi \phi}}}$ and $(0,0,\beta^{\phi})$ refers to the shift vector $\beta^{\phi}=\frac{\sqrt{g_{\phi\phi}} \omega^{\phi}}{\alpha}$. $\omega^{\phi}=-{g_{\phi t }}/{g_{\phi \phi}}$ is the angular velocity associated with frame dragging.

By examining accelerated and decelerated plasma energetics, we examine the viability of magnetic reconnection process for the axially symmetric MRNBH, therefore, it is important to derive the expressions of plasma energies. Here, we use one-fluid approximation for plasma in our case. Within this approximation, the reconnecting plasma surrounding the BH is modeled as a single effective fluid, without separately describing the electron and ion components. Under this condition, the plasma energy-momentum tensor expression becomes as follows:
\begin{eqnarray}
    T^{\mu \nu}=pg^{\mu \nu}+{\mathit{w}}U^{\mu}U^{\nu}+F^{\mu}_{\delta}F^{\nu \delta}-\frac{1}{4}g^{\mu \nu}F^{\rho \delta}F_{\rho \delta}\, , 
\end{eqnarray}
with $p$-plasma pressure component, $\omega$-enthalpy density, $U^\mu$-four-velocity, $F^{\mu\nu}$-electromagnetic field tensor. The enthalpy density can be written as
\begin{eqnarray}
\omega = \rho c^2 + \frac{\Gamma}{\Gamma - 1} p\, .
\end{eqnarray}
where $\rho$ is the proper mass density of the bulk plasma, $p$ refers to the total thermodynamic pressure of the plasma treated as a single fluid, $\Gamma$ is the adiabatic index. For further analysis, we take the plasma as relativistically hot and derive plasma energy measured by an observer at infinity. 

For a stationary spacetime, invariance under time translations is described by a timelike Killing vector field, which can be written as \[\xi_{(t)}^{\mu} =\left(\frac{\partial}{\partial t}\right)^{\mu}= \delta_{t}^{\mu}.\] To describe matter or field using the energy-momentum tensor $T^{\mu \nu}$, the conserved energy with this symmetry can be written as 
\begin{eqnarray}
\mathcal{E}^{\mu} = -T^{\mu\nu}\xi_{\nu}\, .
\end{eqnarray}
The Killing vector can be lowered with the spacetime metric $g_{\mu\nu}$ as:
\begin{eqnarray}
\xi_{\nu} = g_{\nu\lambda}\xi^{\lambda}
= g_{\nu\lambda}\delta^{\lambda}_{t}
= g_{\nu 0}\, ,
\end{eqnarray}
so the conserved energy takes the form
\begin{eqnarray}
\mathcal{E}^{\mu} = -T^{\mu\nu} g_{\nu 0}.
\end{eqnarray}
The energy density is derived by considering a constant-$t$ hypersurface and selecting the time-like component, $\mu=0$,
\begin{eqnarray}
\mathcal{E}^{0} = -T^{0\nu} g_{\nu 0}\, .
\end{eqnarray}
This quantity is measured with respect to coordinate time, not a proper time. Using the transformation that we have therefore introduced above
$d\hat{t}=\alpha dt$, the energy density at infinity takes the form as
\begin{eqnarray}
e^{\infty} = \alpha \mathcal{E}^{0}
= -\alpha T^{0\nu} g_{\nu 0}\, .
\end{eqnarray}
Using the energy-momentum tensor symmetry property $T^{\mu\nu}=T^{\nu\mu}$, we can write $T^{0\nu}=T^{\nu0}$, and by relabeling the index $\nu\to\mu$, we can obtain
\begin{eqnarray}
e^{\infty} = -\alpha g_{\mu 0} T^{\mu 0}\, .
\end{eqnarray}
This refers to the plasma energy density measured by an observer at infinity. The energy density at infinity can be written using azimuthal component of the momentum density $\hat{P}^{\phi}$ and the total energy density $\hat{e}$ as
\begin{eqnarray}
e^{\infty} = \alpha( \hat{e}+ \beta^{\phi}\hat{P}^{\phi})\, ,
\end{eqnarray}
the components can be written as
\begin{eqnarray}
    \hat{e}={\mathit{w}}\hat{\gamma}^2-p+\frac{\hat{B}^2+\hat{E}^2}{2}\, ,\\
   \hat{P}^{\phi}={\mathit{w}}\hat{\gamma}^2\hat{v}^{\phi}+(\hat{B} \times \hat{E})^{\phi}\, ,
\end{eqnarray}
where $\hat{v}^{\phi}$ is $\phi$-component of plasma velocity at the ZAMO frame and $\hat{\gamma} = \hat{U}^{0}
= \frac{1}{\sqrt{1-\sum_{i=1}^{3}\left(\hat{v}^{i}\right)^{2}}}$ is the relativistic gamma factor. We note that the total energy density consists of contributions from both the electromagnetic and hydrodynamic components: $e^{\infty}=e_{hyd}^{\infty}+e_{em}^{\infty}$. Taking electric ($\hat{E}^i=\hat{F}^{i0}$) and magnetic ($\hat{B}^i=\epsilon^{ijk}\hat{F}_{jk}/2$) field components into account, the energy density components in the ZAMO fame can be written as
\begin{figure*} 
\begin{center} \begin{tabular}{c c} 
\includegraphics[scale=0.51]{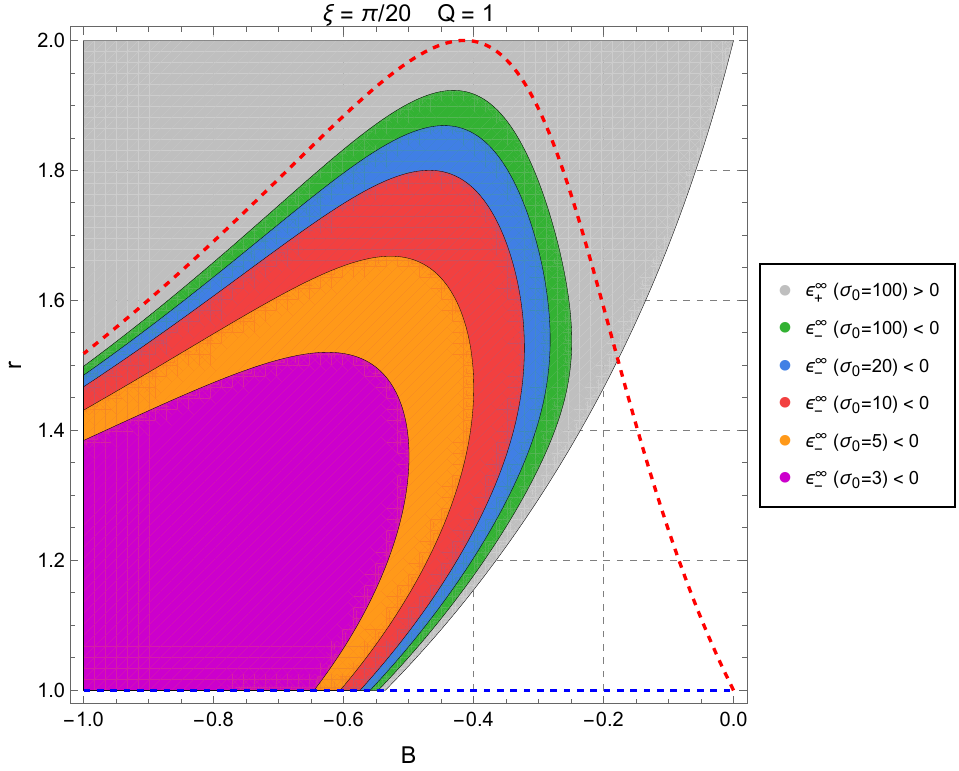}\hspace{1cm} 
\includegraphics[scale=0.51]{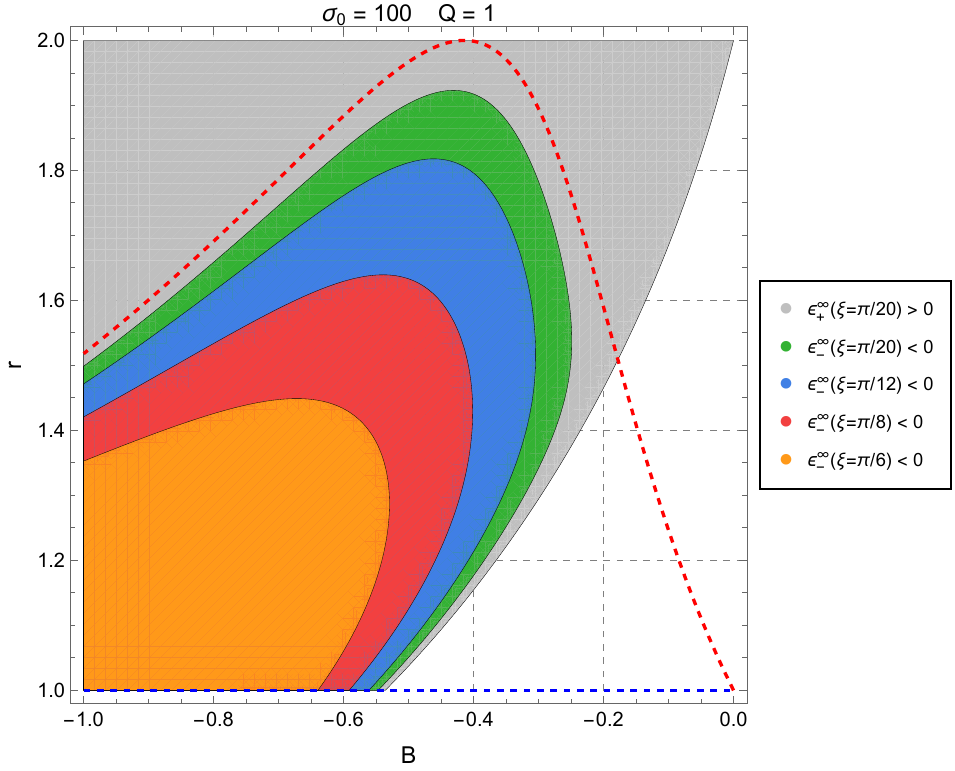}\hspace{1cm}
\end{tabular} \caption{\label{Fig:region2} {The allowed energy-extraction regions for accelerated ($\epsilon_+^{\infty}>0$) and decelerated ($\epsilon_-^{\infty}<0$) plasma states are plotted in the $B-r$ plane for varying plasma magnetization parameter at the fixed orientation angle parameter (Left panel) and for varying orientation angle at the fixed plasma magnetization parameter (Right panel). Noe that red and blue dashed lines, respectively, refer to the boundary of the stationary-limit surface (i.e., ergoregion) and outer event horizon. Here, extremal value of the BH charge is considered and set to unity. }}
\end{center}
\end{figure*}
\begin{eqnarray}
e_{\rm hyd}^{\infty}
&=&
\alpha \omega\hat\gamma^2-\alpha p
+
\alpha \beta^{\phi} w \hat{\gamma}^{2} \hat{v}^{\phi},
\\
e_{\rm em}^{\infty}
&=&
\alpha \hat B^2/2+\alpha \hat E^2/2
+
\alpha \beta^{\phi}
\left(
\hat{\mathbf{B}} \times \hat{\mathbf{E}}
\right)^{\phi}.
\end{eqnarray}
It must be emphasized that the hydrodynamic component of the energy density contributes to the total energy density significantly and that the contribution of the electromagnetic one can be considered to be much more negligible. Taking this into account for the bulk plasma one can write $e^{\infty}$ as
\begin{eqnarray}
e^{\infty} = e_{\rm hyd}^{\infty}
= \alpha \left[
w \hat{\gamma}
\left(1+\beta^{\phi}\hat{v}^{\phi}\right)
-\frac{p}{\hat{\gamma}}
\right]\, .
\end{eqnarray}
To analyze equatorial circular motion of a corotating plasma we can write Keplerian angular velocity as
\begin{widetext}
\begin{equation}
\begin{aligned}
\Omega_K = -\Bigg[
&-\frac{
4BQ\left(16-24B^2r^2+B^4r^4\right)
}{
16r\left(-4+B^2r^2\right)
}
\\[2mm]
&-\frac{
\left[
64+B^2r^2\left(48+12B^2r^2+B^4r^4\right)
\right]
\sqrt{
\begin{aligned}
&-2-\frac{Q^2}{r^2}+\frac{1}{r}
+\frac{32\left[-2+B^2\left(Q^2-r\right)\right]}
{\left(4+B^2r^2\right)^2}
\\
&\hspace{2cm}
+\frac{24+2B^2r}
{4+B^2r^2}
\end{aligned}
}
}{
16r\left(-4+B^2r^2\right)
}
\Bigg],
\end{aligned}
\label{eq:OmegaK}
\end{equation}
\end{widetext}
and Keplerian velocity of a rotating plasma is expressed as
\begin{eqnarray}
\hat{v}_{K}
=
\frac{d\hat{x}^{\phi}}{d\hat{t}}
=
\frac{\sqrt{g_{\phi\phi}}}{\alpha}\Omega_{K}
-
\beta^{\phi}.
\end{eqnarray}
\begin{figure*}
    \centering
    \includegraphics[scale=0.92]{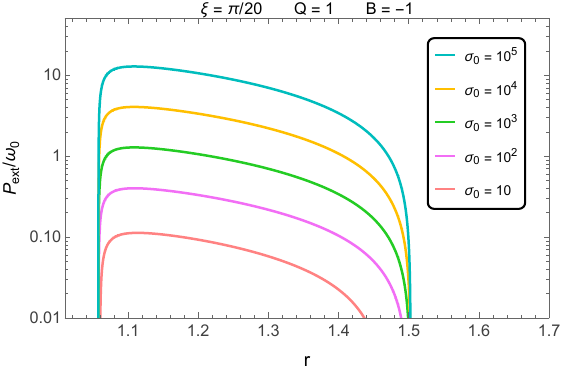}
    \includegraphics[scale=0.92]{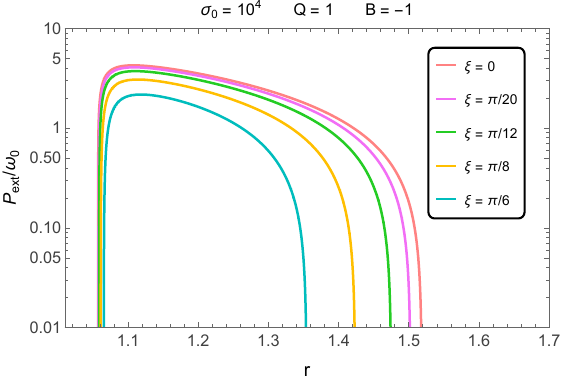}
    \caption{\label{fig:power} {The power which can be extracted from the axisymmetric MRNBH via the magnetic reconnection mechanism is provided as a function of reconnection occuring location $r$. In the left panel, the relation is given for fixed orientation angle value $\xi=\pi/20$ and altering quantities of reconnection magnetization parameter ($10-10^5$). In the right panel, the relation is provided for fixed magnetization factor $\sigma_0=10^4$ and different orientation angle values ($0-\pi/6$). Here BH charge and mass are both taken as 1 and the magnetic field parameter is considered to be $B=-1$.  }}
\end{figure*}
In the above equations, we have used the vector transformation Boyer-Lindquist $\to$ ZAMO for $\psi$ vector as
\begin{eqnarray}
\hat{\psi}^{0} &=& \alpha \psi^{0}, 
\qquad
\hat{\psi}^{i}
=
\sqrt{g_{ii}}\,\psi^{i}
-
\alpha \beta^{i}\psi^{0},
\\
\hat{\psi}_{0}
&=&
\frac{\psi_{0}}{\alpha}
+
\sum_{i=1}^{3}
\frac{\beta_{i}}{\sqrt{g_{ii}}}\,\psi_{i},
\qquad
\hat{\psi}_{i}
=
\frac{\psi_{i}}{\sqrt{g_{ii}}}.
\end{eqnarray}
Using all expressions and considerations above, we can write accelerated and decelerated plasma energies at infinity per unit enthalpy $e^\infty/\omega$ as
\begin{widetext}
\begin{equation}
\epsilon_\pm^\infty =
\frac{\sqrt{\mathcal{A}\mathcal{Z}}}{\sqrt{1-(\mathcal{T}+\mathcal{U})^2}}
\left[
\left(1-\mathcal{T}(\mathcal{T}+\mathcal{U})\right)\sqrt{1+\sigma_0}
\pm \mathcal{U}\sqrt{\sigma_0}\cos\xi
-\frac{
\left[1-(\mathcal{T}+\mathcal{U})^2\right]
\left[
\sqrt{1+\sigma_0}
\mp(\mathcal{T}+\mathcal{U})\sqrt{\sigma_0}\cos\xi
\right]
}{
4\left[
1+\sigma_0-(\mathcal{T}+\mathcal{U})^2\sigma_0\cos^2\xi
\right]
}
\right].
\label{Eq:essential}
\end{equation}
\end{widetext}
Here, $\sigma_0 = B_0^2/\omega_0$ refers to the magnetization parameter of the plasma in the reconnection layer, while $\xi$ denotes the orientation angle between the magnetic field lines and the azimuthal direction in the equatorial plane. Note that the explicit expressions for the quantities appearing in Eq. (\ref{Eq:essential}) are provided in Appendix~\ref{app:supporting}. We have taken the plasma as relativistically hot, so $\omega=4p$ must be satisfied. The positive and negative signed energy at infinity per unit enthalpy expressions define accelerated and decelerated plasma state energies, respectively. For energy extraction, the following conditions $\epsilon_-^\infty<0$ 
and $\Delta \epsilon_{+}^{\infty}
=
\epsilon_{+}
-
\left(
1-\frac{\Gamma}{\Gamma-1}\frac{p}{w}
\right)
=
\epsilon_{+}^{\infty}
>0$ must be satisfied for the plasma. The value $\Gamma=4/3$ can be taken for the relativistically hot plasma case. 

We then turn to analyze how the relevant reconnection parameters affect the conditions and energy extraction efficiency. It must be noted that for a positively charged MRNBH, energy extraction can be done for negative values of the magnetic field parameter $B$ within the convention used here. The sign of $B$ specifies the orientation of the magnetic field rather than its strength, so the sign of the angular velocity of frame-dragging is opposite to that of the product $QB$. Therefore, this can lead to $\omega^\phi>0$, while the same signs give rise to  $\omega^\phi<0$. For $Q>0$ and $B<0$, we obtain $\omega^\phi>0$. In this configuration, the induced frame dragging is in the same direction as the orbital motion of the plasma, making it easier for the decelerated reconnection outflow to reach a negative-energy state, i.e., $\epsilon_{-}^{\infty}<0$. In Fig. \ref{Fig:energyatinfsigma}, we show the energies at infinity of the accelerated and decelerated plasma, \(\epsilon_+^\infty\) and \(\epsilon_-^\infty\), as functions of the magnetization parameter \(\sigma_0\) for axially symmetric MRNBH for varying the magnetic field parameter $B$. As shown in Fig. \ref{Fig:energyatinfsigma}, the accelerated plasma energy increases noticeably as compared to the decelerated plasma energy, with increasing both the magnetization $\sigma_0$ and the magnetic field $B$ parameters.  
\begin{figure*}
    \centering
    \includegraphics[scale=0.7]{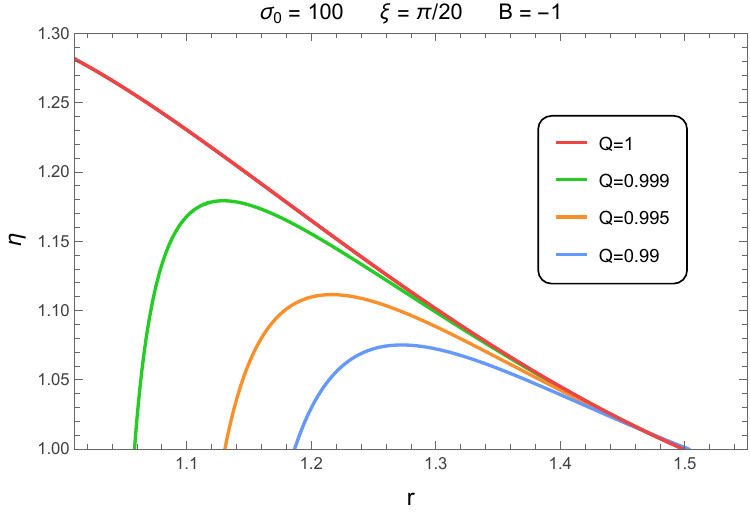}
    \includegraphics[scale=0.7]{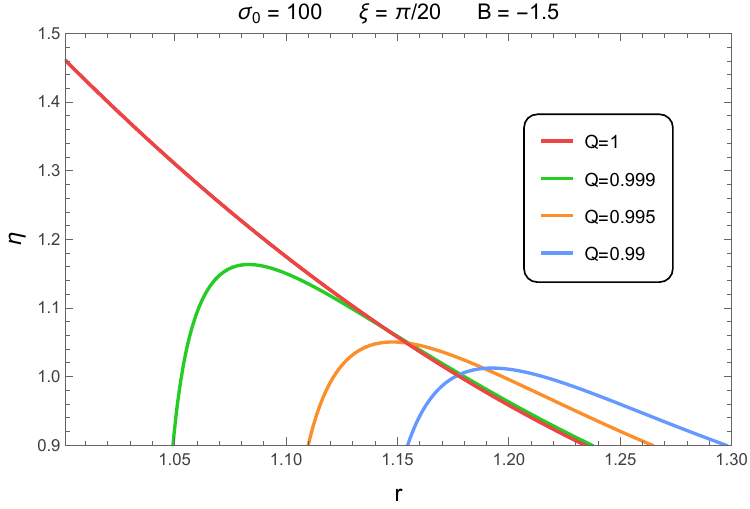}
    \includegraphics[scale=0.7]{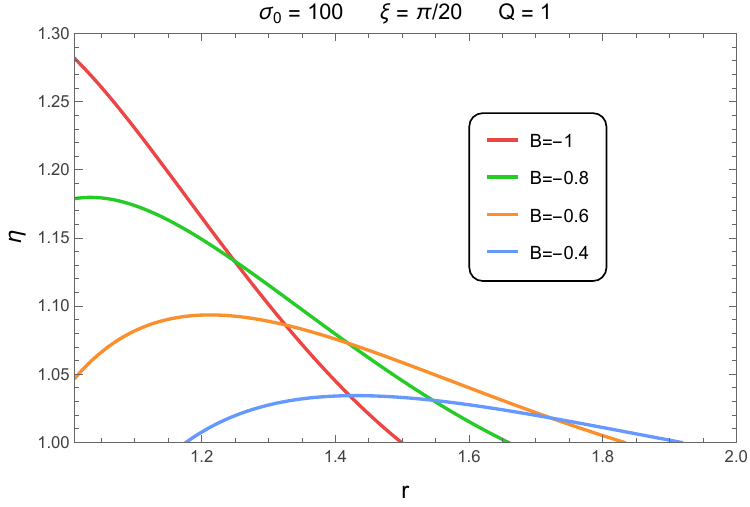}
    \caption{\label{fig:efficiency} Energy-extraction efficiency of the magnetic reconnection mechanism as a function of the reconnection radius for the axially symmetric MRNBH. The top-left and right panels show the results for different values of the BH charge $Q$ at fixed magnetic field parameters $B=-1$ and $B=-1.5$, respectively, while the bottom panel shows the dependence on $B$ for fixed $Q=1$. Here, we set $M=1$, $\sigma_0=100$, and $\xi=\pi/20$.} 
\end{figure*} 

As already stated, for energy extraction to be possible, $\epsilon_-^\infty<0$ and $\epsilon_+^\infty>0$ must be always satisfied for accelerated/decelerated plasma energies. In Fig. \ref{fig:region1}, we analyze the regions satisfying energy extraction conditions in the $Q-r$ plane. In the left of Fig. \ref{fig:region1}, we present the energy extraction regions allowed for varying reconnection magnetization parameter ($\sigma_0=3-100$) at the fixed reconnecting orientation angle $\pi/20$, while the right panel shows the same for varying reconnecting orientation angle value ($\xi=\pi/20-\pi/6$) at the fixed reconnection magnetization parameter $\sigma_0=100$ and. Note that red and blue dashed lines, respectively, represent boundaries of ergosphere and outer event horizon. As can be seen from Fig. \ref{fig:region1}, a higher reconnection magnetization parameter and a lower orientation angle cause to broader allowed energy extraction regions, thus resulting in a higher efficiency of energy extraction. In addition, it is worth noting that higher energy extraction efficiency are obtained as the black hole charge approaches its extremal value. Based on the mechanism, energy extraction is possible only within the ergoregion, i.e., in phases between outer event horizon and the stationary-limit surface. In our analysis, all parameter regions that allow energy extraction lie between these two boundaries, indicating by red and blue dashed lines. This confirms that the extraction conditions are satisfied throughout the allowed region considered here in our calculations.

In Fig. \ref{Fig:region2}, we demonstrate the allowed regions for energy extraction in the $B-r$ plane for an extremal BH case for varying reconnection orientation angle and magnetization parameters, respectively. From Fig. \ref{Fig:region2}, we can observe the same trend, i.e. increasing the magnetization parameter and decreasing the orientation angle can broaden the parameter region in which energy extraction is possible, consequently enhancing the extraction efficiency. 
As shown in Fig. \ref{Fig:region2}, the allowed energy-extraction regions are mainly associated with negative values of the magnetic field parameter $B$. As the magnetic field increases for $B<0$, these regions broaden and extend over a larger radial range $r$, whereas they become gradually restricted as $B$ approaches zero. This behavior indicates that a magnetic field with the opposite orientation relative to the BH charge favors magnetic reconnection-driven energy extraction. The red and blue dashed lines represent the stationary-limit surface and the outer event horizon, respectively. All allowed extraction regions lie between these two boundaries, which confirms that energy extraction occurs within the ergoregion.
\begin{figure}
    \centering
    \includegraphics[scale=1.0]{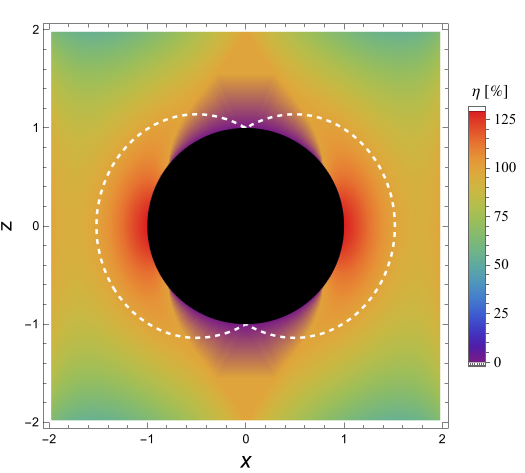}
    \caption{\label{fig:efficiencydistribution} {Distribution of the energy-extraction efficiency via magnetic reconnection for the axially symmetric MRNBH in the $X-Z$ plane. The boundaries of the ergoregion are shown by the white dashed curves. Here, the BH and magnetic reconnection parameters are chosen suitable for maximum energy extraction conditions ($M=1, Q=1, B=-1, \sigma_0=100$ and $\xi=\pi/20$).} }
\end{figure}

\section{\label{Sec:RN-Power-MR} Power and Efficiency of Energy Extraction via Magnetic Reconnection}

In this section, we investigate the power and efficiency of energy extraction through magnetic reconnection in the axially symmetric MRNBH spacetime under the conditions discussed above. The energy carried by the infalling (decelerated) plasma plays a central role in determining the extractable power. In particular, the power, $P_{ext}=-\epsilon_{-}^{\infty}{\mathit{w}}_0 A_{in} U_{in}$, extracted from the black hole through magnetic reconnection is directly related to the negative energy of the decelerated plasma and can be written as 
\begin{widetext}
\begin{equation}
P_{ext} =
-\frac{ \omega_0 A_{in} U_{in} \sqrt{\mathcal{A}\mathcal{Z}}}{\sqrt{1-(\mathcal{T}+\mathcal{U})^2}}
\left[
\left(1-\mathcal{T}(\mathcal{T}+\mathcal{U})\right)\sqrt{1+\sigma_0}
- \mathcal{U}\sqrt{\sigma_0}\cos\xi
-\frac{
\left[1-(\mathcal{T}+\mathcal{U})^2\right]
\left[
\sqrt{1+\sigma_0}
+(\mathcal{T}+\mathcal{U})\sqrt{\sigma_0}\cos\xi
\right]
}{
4\left[
1+\sigma_0-(\mathcal{T}+\mathcal{U})^2\sigma_0\cos^2\xi
\right]
}
\right]\, ,
\label{Eq:power}
\end{equation}
\end{widetext}
where $\epsilon_{-}^{\infty}$ is the infalling (decelerated) plasma energy given by Eq. (\ref{Eq:essential}), $\omega_0$ is the density of enthalpy. $A_{in}$ and $U_{in}$, respectively, denote effective cross-sectional area of the infalling plasma and the dependent regime of the reconnection mechanism. We consider that energy extraction occurs in the collisionless regime, i.e. $U_{in}={\cal O}(10^{-1})$ is chosen for further analysis. Additionally, the effective cross-sectional area of the inflowing plasma is approximated by $A_{\rm in} \sim \left(r_{\rm st}^{2}-r_{\rm H}^{2}\right)$.  

Given the importance of Eq.~(\ref{Eq:power}), we examine the relation between the extracted power per unit enthalpy density $P_{ext}/\omega_0$ in the radial location $r$. In Fig. \ref{fig:power}, we represent this relation $P_{ext}/\omega_0$. In the left panel of Fig.~\ref{fig:power}, we show the relation mentioned above for the orientation angle $\xi=\pi/20$ and for the reconnection magnetization parameter $\sigma_0$ varying from 10 to $10^5$, while in the right panel we show it for the magnetization parameter $\sigma_0=10^4$ and for the orientation angle $\xi$ varying from 0 to $\pi/6$. As seen in Fig~\ref{fig:power}, larger reconnection magnetization parameter and smaller orientation angles lead to higher extracted power within the ergoregion. The radial dependence further shows that magnetic reconnection is most effective close to the event horizon, where the conditions for producing negative-energy plasma are strongest. As the reconnection region moves outward, these conditions weaken, causing the extracted power to gradually decrease and eventually vanish.

We further analyze and assess the energy extraction efficiency of the magnetic reconnection process, which can be defined as 
\begin{widetext}
\begin{equation}
\eta \equiv
\frac{\epsilon_+^\infty}
{\epsilon_+^\infty+\epsilon_-^\infty}
=
\frac{
\left[1-\mathcal{T}(\mathcal{T}+\mathcal{U})\right]
\sqrt{1+\sigma_0}
+\mathcal{U}\sqrt{\sigma_0}\cos\xi
-\frac{
\left[1-(\mathcal{T}+\mathcal{U})^2\right]
\left[
\sqrt{1+\sigma_0}
-(\mathcal{T}+\mathcal{U})\sqrt{\sigma_0}\cos\xi
\right]
}{
4\left[
1+\sigma_0-(\mathcal{T}+\mathcal{U})^2
\sigma_0\cos^2\xi
\right]
}
}{
2\left[1-\mathcal{T}(\mathcal{T}+\mathcal{U})\right]
\sqrt{1+\sigma_0}
-\frac{
\left[1-(\mathcal{T}+\mathcal{U})^2\right]
\sqrt{1+\sigma_0}
}{
2\left[
1+\sigma_0-(\mathcal{T}+\mathcal{U})^2
\sigma_0\cos^2\xi
\right]
}
}\, ,
\label{Eq:efficiency}
\end{equation}
\end{widetext}
where $\epsilon_{+}^{\infty}$ and $\epsilon_{-}^{\infty}$ are considered the energies of the accelerated and decelerated plasma components at infinity (see Eq. \ref{Eq:essential}). For the decelerated plasma energy, $\epsilon_{-}^{\infty}<0$ leads to the efficiency that becomes $\eta>1$, indicating that energy is extracted from the BH. In Fig. \ref{fig:efficiency}, we show the dependence of energy extraction efficiency on the radial location of the mechanism for different values of BH charge $Q$ (top-left and right panels) and magnetic field parameter $B$ (bottom panel) for the axially symmetric MRNBH. As $Q$ approaches its maximum values, the efficiency of energy extraction by magnetic reconnection increases significantly in the ergoregion. With an increase in the value of $Q$, the efficiency initially increases with $r$ - the radial location of the reconnection, and reaches its maximum. As can be seen from the top-right panel of Fig.~\ref{fig:efficiency}, increasing $Q$ leads to a higher maximum energy extraction efficiency and shifts the corresponding peak toward smaller $r$. It is evident from the right panel of Fig. \ref{fig:efficiency} that the energy efficiency at a sufficiently larger magnetic field $B$ becomes larger and reaches a value comparable to an extremal Kerr BH where it always increases as the rotation parameter grows \cite{Comisso21}. It turns out that the magnetic field parameter enhances the energy extraction capability. This enhancement in the MRNBH is associated with the interaction between the BH charge and the magnetic field. In addition, we demonstrate that increasing the magnetic field eventually increases the magnetic reconnection-driven energy extraction, particularly near the horizon, as seen in the bottom panel of Fig. \ref{fig:efficiency}. Therefore, magnetic reconnection-driven energy extraction becomes more efficient at sufficiently large magnetic field. For the extremal MRNBH case, the energy extraction efficiency reaches nearly $130\%$ for $B=-1$. However, within the parameter ranges considered here, a large magnetic field can allow the efficiency to reach the extremal Kerr value. 
\begin{figure}
    \centering
    \includegraphics[scale=1.0]{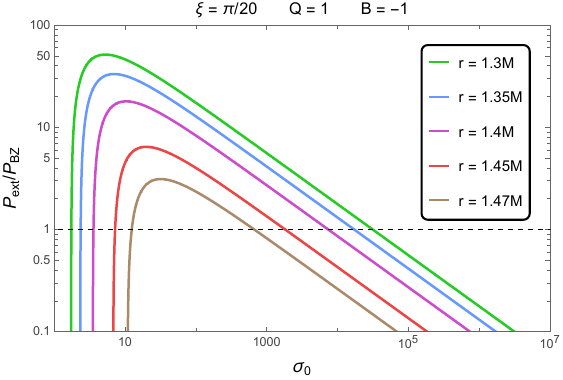}
    \caption{\label{fig:efficiencyratio} Ratio of the power extracted via two different energy extraction mechanisms: magnetic reconnection and the BZ, $P_{\rm ext}/P_{\rm BZ}$, as a function of the reconnection magnetization factor $\sigma_0$ for different values of the radial location of the process. The horizontal dashed line ($P_{\rm MR}/P_{\rm BZ}=1$) represents the equal power extraction of the two mechanisms. Here, $M$ is set to unity.}
\end{figure}
 
The behavior shown in Fig.~\ref{fig:efficiency} is also illustrated in Fig.~\ref{fig:efficiencydistribution}, which shows the spatial distribution of the magnetic-reconnection efficiency in the $X-Z$ plane of the axisymmetric MRNBH spacetime. The efficiency for the fixed magnetic field $B=-1$ reaches values close to $130\%$ within the ergoregion, as seen in Fig.~\ref{fig:efficiencydistribution}. In this region, $\eta>100\%$ indicates the extraction of energy from the black hole, enabled by the formation of negative-energy plasma states. Outside the ergoregion, the efficiency falls below $100\%$ and decreases further with increasing distance from the black hole. Thus, magnetic reconnection-driven energy extraction is confined to the ergoregion, where negative energy plasma states can exist. 
\begin{figure*}
    \centering
    \includegraphics[scale=0.57]{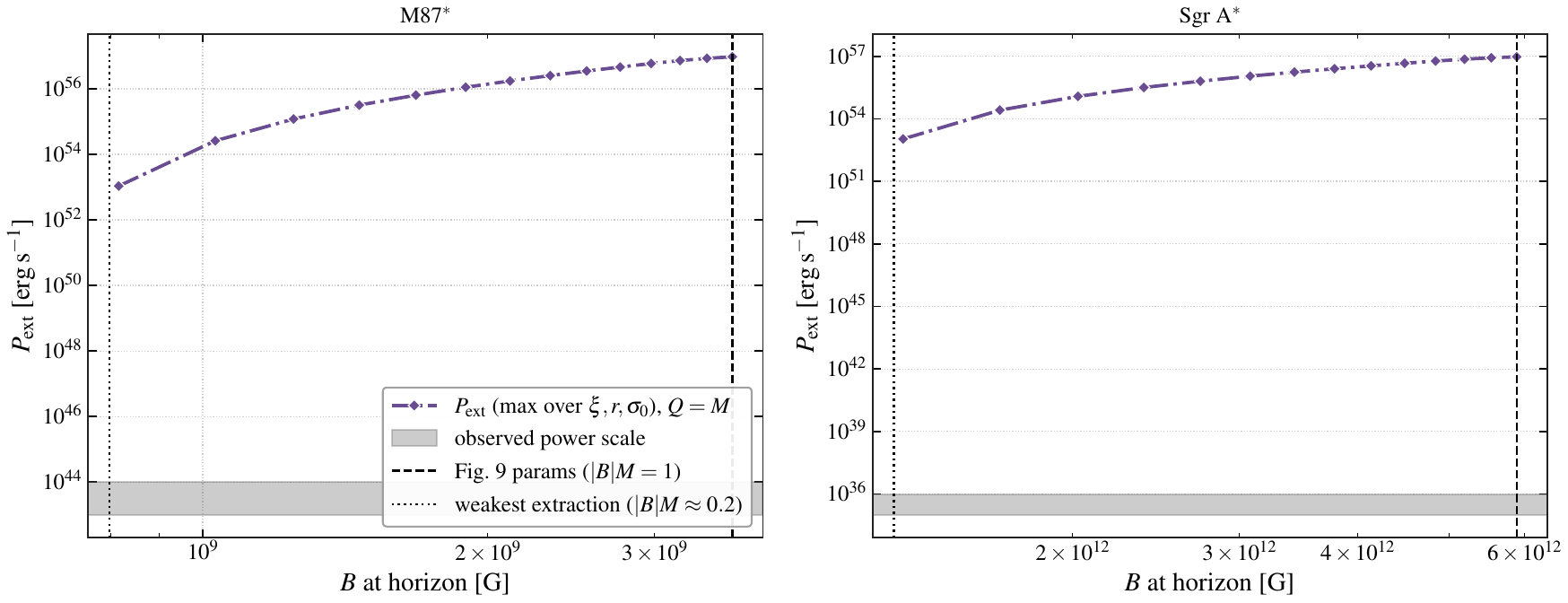}
    \includegraphics[scale=0.57]{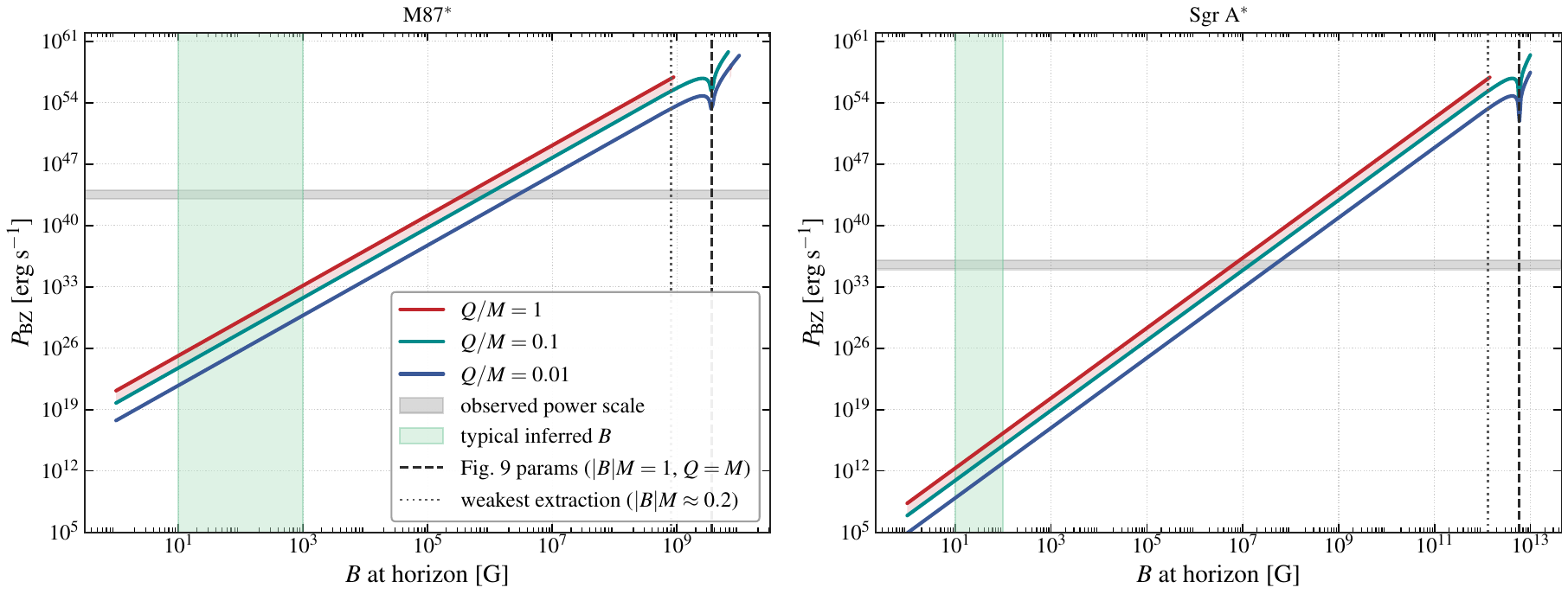}
  \caption{Power output of the axially symmetric MRNBH and its comparison with observations.
\textit{Upper panels:} maximum reconnection-extraction power $P_{\mathrm{ext}}$ (indigo dash-dotted curve with diamond markers), computed from Eq.~\eqref{Eq:essential} with $w=B_0^2/(4\pi\sigma_0)$ and $B_0=B$, maximized over $\xi$, $r$ and $\sigma_0$; at fixed $|B|M$ it is independent of the BH mass. The left and right columns correspond to M87* ($M=6.5\times10^{9}\,M_\odot$) and Sgr~A* ($M=4.0\times10^{6}\,M_\odot$), respectively.
\textit{Lower panels:} BZ power as a function of the magnetic-field strength at the horizon, shown for $Q/M=1$ (crimson), $0.1$ (teal) and $0.01$ (slate blue), with the horizon angular velocity $\Omega_H$. For $Q/M=1$ the shaded crimson band spans $\Omega_H=N(r_H)$ down to $N(r_H)/4$; the Kerr-fit correction of \cite{Tchekhovskoy10ApJ}
remains valid for $\Omega_H\lesssim0.5$, and curves are truncated where this bound is exceeded.
In both rows, the horizontal grey band marks the observed power scale ($P\sim10^{43}$--$10^{44}\,\mathrm{erg\,s^{-1}}$ for M87* and $\sim10^{35}$--$10^{36}\,\mathrm{erg\,s^{-1}}$ for Sgr~A*), while the vertical light-green band marks the range of horizon magnetic fields typically inferred from synchrotron and Faraday-rotation modelling. The dashed vertical line marks the reference parameters $|B|M=1$, $Q=M$ used in Fig.~~\ref{fig:efficiencyratio}; the dotted vertical line marks the weak-field extraction threshold $|B|M\approx0.2$.}
  \label{fig:power_vs_obs}
\end{figure*}

We now compare the energy-extraction capabilities of magnetic reconnection and the Blandford--Znajek (BZ) mechanism under the adopted approximations. In the standard BZ process, black hole rotational energy is extracted electromagnetically through magnetic fields interacting with frame dragging near the horizon. In the axisymmetric MRNBH spacetime considered here, however, frame dragging is not generated by black hole rotation but is induced by the coupling between the black hole charge and the magnetic field. It is therefore useful to examine how this electromagnetically induced frame dragging affects the relative performance of the two extraction mechanisms. The ratio between the magnetic-reconnection power $P_{\rm ext}$ and the BZ power $P_{\rm BZ}$ can be written as \cite{Blandford1977,Tchekhovskoy10ApJ,Comisso21}
\begin{widetext}
\begin{eqnarray}
\frac{P_{ext}}{P_{BZ}} &=&
-\frac{ A_{in} U_{in} \sqrt{\mathcal{A}\mathcal{Z}}}{\kappa \Omega_{H}^2r_{H}^4\sigma_0\sin^2{\xi}(1+\chi\Omega_{H}^2+\zeta\Omega_{H}^4) \sqrt{1-(\mathcal{T}+\mathcal{U})^2}}
\bigg[
\left(1-\mathcal{T}(\mathcal{T}+\mathcal{U})\right)\sqrt{1+\sigma_0}
- \mathcal{U}\sqrt{\sigma_0}\cos\xi
\nonumber\\
&&-\frac{
\left(1-(\mathcal{T}+\mathcal{U})^2\right)
\left(
\sqrt{1+\sigma_0}
+(\mathcal{T}+\mathcal{U})\sqrt{\sigma_0}\cos\xi
\right)
}{
4\left[
1+\sigma_0-(\mathcal{T}+\mathcal{U})^2\sigma_0\cos^2\xi
\right]
}\bigg],
\label{Eq:powercomparison}
\end{eqnarray}
\end{widetext}
here $\epsilon_{-}^{\infty}$ is the decelerated plasma energy, $A_{in}$ is considered the cross-sectional area of the infalling plasma and $U_{in}$ refers to the reconnection occurring regime as we stated above, and taken as $U_{in}=0.1$ (for collisionless regime), $A_{in}=(r_{st}^2-r_H^2)$. Additionally, $\kappa$ is the numerical parameter which defines magnetic field geometry near the BH (in our case, we use split monopole geometry, $\kappa\approx0.05$), $\Omega_H$ is the rotation of the horizon induced by the electromagnetic field, $r_H$ - BH event horizon radius, $\chi$ and $\zeta$ are both numerical constants \cite{Tchekhovskoy10ApJ}.

Fig.~\ref{fig:efficiencyratio} compares the power extracted from the MRNBH through magnetic reconnection, $P_{\rm ext}$, with that associated with the BZ mechanism, $P_{\rm BZ}$, for different reconnection radii. Our results show that magnetic reconnection can produce a substantially larger extracted power than the BZ mechanism, particularly when reconnection occurs close to the event horizon. Consequently, the ratio $P_{\rm ext}/P_{\rm BZ}$ reaches its largest values at small radii and decreases as the reconnection region moves outward. The ratio also exhibits a nonmonotonic dependence on plasma magnetization $\sigma_0$. As $\sigma_0$ increases, $P_{\rm ext}/P_{\rm BZ}$ initially grows, reaches a maximum, and then gradually decreases. Thus, magnetic reconnection can dominate over the BZ mechanism within a finite range of plasma magnetization, whereas at sufficiently large $\sigma_0$, the BZ mechanism becomes comparatively more efficient.

\vspace{0.5cm}
\noindent\textit{\textbf{Observational Consequences:}}

To assess the astrophysical relevance of the energy extraction mechanisms, we compare both channels (magnetic reconnection and the field-induced BZ process) with observations (Fig.~\ref{fig:power_vs_obs}).
We evaluate the reconnection power directly from the decelerated-plasma branch (lower sign) of Eq.~\eqref{Eq:essential}, $P_{\mathrm{ext}}=-\epsilon^{\infty}_{-}\,\omega_0\,U_{\mathrm{in}}A_{\mathrm{in}}\,c$, where $\epsilon^{\infty}_{-}$ is the energy at infinity per unit enthalpy and $\omega_0=B_0^2/(4\pi\sigma_0)$ in Gaussian units ($\omega_0=B_0^2/\sigma_0$ in the units used for $\sigma_0$ above). We identify the layer field $B_0$ with the horizon field $B$ and maximize over $\xi$, $r$ and $\sigma_0$ (upper panels). At fixed dimensionless field strength $|B|M$ (with $B$ in geometric units, $G=c=1$), the reconnection power depends on the black hole only through this product, $P_{\mathrm{ext}}\propto(BM)^{2}c^{5}/G$, so M87* and Sgr~A* share the same $P_{\mathrm{ext}}$ at equal $|B|M$ and differ only in the corresponding field strength in gauss, $B_{\mathrm{G}}\propto(|B|M)/M$. 
For $Q=M$, extraction ($\epsilon^{\infty}_{-}<0$) occurs only for $|B|M\gtrsim0.2$, i.e. $B\gtrsim8\times10^{8}\,\mathrm{G}$ for M87* and $B\gtrsim1.3\times10^{12}\,\mathrm{G}$ for Sgr A*, and the maximum power rises from $\sim10^{53}$ to $\sim10^{57}\,\mathrm{erg\,s^{-1}}$ as $|B|M$ increases from 0.2 to 1 (the Fig.~\ref{fig:efficiencyratio} value, $B\approx4\times10^{9}\,\mathrm{G}$ for M87* and $B\approx6\times10^{12}\,\mathrm{G}$ for Sgr A*).
This exceeds the observed power scale by $\sim$9-14 orders of magnitude for M87* and $\sim$17-22 for Sgr A*, and the Eddington luminosity by $\sim$5-9 and $\sim$8-12 orders, respectively.
No extraction was found for $|B|M\lesssim0.2$ in any scanned combination, down to the weakest field we sampled ($|B|M=10^{-3}$, $\approx4\times10^{6}\,\mathrm{G}$ for M87*); the fields inferred from observations lie well below this range, where the ergoregion shrinks further, so we expect $P_{\mathrm{ext}}=0$ there.
For the field-induced BZ channel, using the normalization of Ref.~\cite{Tchekhovskoy10ApJ}, $P_{\mathrm{BZ}}\approx\frac{1}{6}B^2r_H^4\Omega_H^2/c$ with $\Omega_H=N(r_H)$ (lower panels; the shaded band spans $\Omega_H=N(r_H)$ to $N(r_H)/4$), the power at the inferred fields is $\sim10^{24}$--$10^{33}\,\mathrm{erg\,s^{-1}}$ for M87* ($10$--$10^{3}$\,G) and $\sim10^{11}$--$10^{16}\,\mathrm{erg\,s^{-1}}$ for Sgr A* ($10$--$10^{2}$\,G), even for $Q=M$, far below the observed jet-power scale.
Neither channel (magnetic reconnection or the field-induced BZ process) therefore reproduces the observed jet power: at inferred fields both are far too weak, while at the fields where reconnection operates the predicted power greatly exceeds observations.
Because reconnection is intrinsically localized and intermittent. The flares are often considered its natural observational counterpart \cite{Comisso:2020ykg,Ripperda:2021zpn}. Although the powers obtained here exceed the flare luminosities of Sgr~A* \cite{Neilsen:2013sha,Zhang:2017qrv} and M87* \cite{Abramowski:2012,Berge:2006uls} by many orders of magnitude, so this framework does not reproduce them either. To match flares in this framework would require a much smaller effective emitting fraction of the reconnection layer, which we leave for future work. In the literature, reconnection has also been discussed near the base and boundary of jets, where extracted energy can heat and accelerate plasma \cite{Ripperda:2021zpn}, 
while for rotating BHs the BZ process is the standard source of the bulk jet power \cite{Blandford1977,Tchekhovskoy10ApJ}.

\section{Conclusion}
\label{Sec:Conclusion}

In this paper, we applied the magnetic reconnection energy extraction mechanism to the axially symmetric MRNBHs. In this BH spacetime, the combination of the BH's charge $Q$ and the $B$ parameters causes frame-dragging effects and an ergoregion formation even without the spin parameter, unlike a Kerr BH. In the ergoregion, the magnetic reconnection divides the plasma into accelerated and decelerated components. When the required conditions are satisfied, the decelerated component of plasma possesses negative energy and falls into the BH, while the accelerated one goes to infinity having positive energy of the BH. Therefore, the magnetic reconnection mechanism can effectively extract energy from the axially symmetric MRNBH even though the BH itself is non-rotating.

 We showed that the frame dragging reverse the direction at the point where $\Omega_{\rm ZAMO}=0$ as shown by the dashed curve in Fig.~\ref{fig:omega_zamo_ergoregion} and solid red curve in Fig.~\ref{fig:omega_zamo_vs_B}. In Fig.~\ref{fig:omega_zamo_ergoregion}, we also showed that it is possible for the ergoregion to become disconnected from the black hole and extend to large distances when the value of the $B$ parameter increases in the negative direction. However, we also observed that the size of the ergoregion does not increase indefinitely with increasing $|B|$. Instead, it reaches a maximum before decreasing (see more details in \cite{Shaymatov:2022eyz}).
We also established the distinction between the two types of ergoregions, namely geometrical and induced, in view of the recent analysis performed in \cite{Baez:2026ich} to remove the misunderstanding between the two ergoregions. In \cite{Baez:2026ich}, the negative energy region of charged test particles was considered an ergosphere. Nevertheless, the zone in question had a dependency on the charge-to-mass ratio $e=q/m$ of the probe particle via the inequality $g_{tt} + (eA_t)^2 = 0$. Hence, it is important to note that it was not a true geometrical ergosphere. The true geometrical ergoregion was determined exclusively by the causal nature of the stationary Killing vector, which can be written as $g_{tt} > 0$. The difference between the two types of ergospheres was necessary for correctly analyzing energy extraction from the neutral and charged particles in MRNBH spacetimes.

We analyzed the accelerated/decelerated plasma energies at infinity as a function of the reconnection magnetization parameter for varying magnetic field at the fixed BH's $Q$, as seen in Fig.~\ref{Fig:energyatinfsigma}.  
In our analysis, the generated frame-dragging helps the decelerated plasma flow to reach negative energy $\epsilon_-^\infty<0$, allowing energy extraction. We showed that both the magnetic field and the reconnection magnetization significantly influence the plasma energies $\epsilon_{+}^{\infty}$ and $\epsilon_{-}^{\infty}$ (see Fig.~\ref{Fig:energyatinfsigma}).  Further we examined the allowed energy extraction regions for accelerated/decelerated $\epsilon_{+}^{\infty}>0$ and $\epsilon_{-}^{\infty}<0$ as functions of BH charge and magnetic field parameter for different values of the reconnection magnetization $\sigma_0$ and orientation angle $\xi$. We found that the region between outer event horizon and the ergoregion boundary, where magnetic reconnection can occur under the conditions required for this process, leading to larger accelerated/decelerated $\epsilon_{+}^{\infty}$ and $\epsilon_{-}^{\infty}$ as a consequence of the interaction between $Q$ and $B$, $\sigma_0$ and lower orientation angles $\xi$. We found that higher plasma magnetization and smaller outflow orientation angles broadened the allowed energy extraction region and increased the extraction efficiency (see Fig.~\ref{fig:region1}). The efficiency increased further as the BH's $Q$ approached its extremal value. We also found that increasing the magnetic field $|B|$ extends the extraction region over a wider radial range, while this region becomes gradually restricted as $B$ approaches zero (see Fig.~\ref{Fig:region2}). These results highlight all identified extraction regions lie between the outer event horizon and the stationary-limit surface, therefore satisfying the ergoregion requirement of the magnetic reconnection mechanism.

Further we found that the extracted power $P_{ext}/\omega_0$ increases with plasma magnetization and decreases with the reconnection orientation angle. This power originates from the negative-energy decelerated plasma falling into the black hole, which allows the accelerated outflow to carry away extracted energy. The extracted power is therefore largest near the event horizon, where negative-energy states are most strongly favored, and decreases as the reconnection location moves outward, eventually vanishing when the negative-energy condition was no longer satisfied (see Fig.~\ref{fig:power}). Our findings showed that strongly magnetized plasma and small outflow angles favor powerful energy extraction near the horizon.

Furthermore, we assessed the energy extraction efficiency of the magnetic reconnection mechanism in the axially symmetric MRNBH. We found that the BH's $Q$ and magnetic field $B$ strongly influence the efficiency of magnetic reconnection-driven energy extraction (see Fig.~\ref{fig:efficiency}). For a fixed $Q$, the efficiency initially increases with the reconnection radius and reaches a maximum within the ergoregion. Increasing $Q$ toward its extremal value raises this maximum and shifts its location toward smaller radii. Increasing $B$ further enhances the efficiency in the near-horizon region around an extremal MRNBH, and it reaches nearly $130\%$ at $B=-1$ (see in the top-left panel of Fig.~\ref{fig:efficiency} and Fig.~\ref{fig:efficiencydistribution}). For a sufficiently large magnetic field, the energy efficiency becomes comparable to an extremal Kerr BH case \cite{Comisso21}, as shown in the top-right panel of Fig.~\ref{fig:efficiency}. In addition, we compared the power extracted from the MRNBH through magnetic reconnection, $P_{\rm ext}$, with that associated with the BZ mechanism, $P_{\rm BZ}$, for different reconnection radii (see Fig.~\ref{fig:efficiencyratio}). Magnetic reconnection can extract substantially more power from the MRNBH than the BZ mechanism. We found that it also depends nonmonotonically on magnetization parameter, i.e. the ratio $P_{\rm ext}/P_{\rm BZ}$ rises with $\sigma_0$ at small reconnection radii. Magnetic reconnection therefore dominates only within a finite range of magnetization, while the BZ mechanism becomes more efficient at sufficiently large $\sigma_0$.

Finally, we assessed the astrophysical relevance of the magnetic reconnection and BZ processes by comparing them with the observed power scales of M87* and Sgr~A*. The reconnection power $P_{\mathrm{ext}}$ was analyzed numerically via $\epsilon_{-}^{\infty}$, with the layer magnetic field $B_0$ equal to the horizon magnetic field $B$, and maximized with respect to the direction of the outflow $\xi$, the position of reconnection $r$ and plasma magnetization $\sigma_0$, see Fig.~\ref{fig:power_vs_obs}. 
It is important to note that at the fixed $|B|M$,  $P_{\mathrm{ext}}$ is independent of the BH mass parameter $M$, so the two sources M87* and Sgr~A* differ only in the corresponding $B$ strength in Gauss. We found that for $Q=M$, energy extraction requires $|B|M\gtrsim0.2$, corresponding to $B\gtrsim8\times10^{8}\,\mathrm{G}$ for M87* and $B\gtrsim1.3\times10^{12}\,\mathrm{G}$ for Sgr~A*. The $P_{ext}$ grows from $\sim10^{53}$ to $\sim10^{57}\,\mathrm{erg\,s^{-1}}$ as $|B|M$ increases from $0.2$ to $1$. This is $9$-$14$ orders of magnitude higher than the observed power scale for M87* and $17$-$22$ for Sgr~A*, and $5$-$9$ and $8$-$12$ orders above Eddington, respectively. 
For smaller $B$ fields, such as the ones deduced observationally ($\sim10$--$10^{3}$\,G), we interestingly find no energy extraction region appears to exist in the scanned parameter space. For comparison, we plotted the $P_{BZ}$ as a function of $B$, estimated with the normalization of Ref.~\cite{Tchekhovskoy10ApJ}. We noticed $P_{BZ}$ is $\sim10^{24}$--$10^{33}\,\mathrm{erg\,s^{-1}}$ for M87* and $\sim10^{11}$--$10^{16}\,\mathrm{erg\,s^{-1}}$ for Sgr~A* at the inferred fields, even for the maximal charge $Q=M$, many orders of magnitude below the observed jet power (see the red curve within the light green band of the lower panel of Fig.~\ref{fig:power_vs_obs}). Neither channel (magnetic reconnection or the field-induced BZ process) therefore reproduces the observed jet power in the light-green region, as they are far below the observed scale as shown by the grey color in Fig.~\ref{fig:power_vs_obs}. However, reconnection greatly overshoots it where it operates, while at the inferred fields neither mechanism approaches it.

In this observational context, it is worth pointing out that magnetic reconnection is necessarily localized and intermittent process, and hence flares are the natural observable counterpart of such events. Therefore, finding an appropriate match for this scenario would require a significantly reduced effective emitting fraction of reconnected $B$ field, compared to that assumed in this paper, and is beyond the scope of the current work. However, in any case, magnetic reconnection might also provide a contribution at the base of the jet, whereas the BZ mechanism provides the dominant power for the jet. Our calculations are of order of magnitude, assuming the aforementioned assumptions (particularly $B_0=B$, the parameter space considered in the scans, and the $\Omega_H$ range used in the estimation of BZ luminosity). Hence, the comparison imposes the constrains on the product of MRNBH's $Q$ and $B$ parameters that a charged magnetized BZ mechanism would need in order to power the observed emission.

\acknowledgments

PS would like to acknowledge financial support from the Anusandhan National Research Foundation (ANRF), New Delhi, under grant number CRG/2023/008980.

\appendix
\section{Supporting expressions}
\label{app:supporting}

Here, we provide the explicit expressions for the quantities
appearing in the plasma energies, energy extraction efficiency, extracted power and power ratios 
in Eqs. \cref{Eq:essential,Eq:power,Eq:efficiency,Eq:powercomparison}

\appendix
\begin{equation}
\mathcal{A} = 1+\frac{B^2r^2}{2}+\frac{B^4r^4}{16},
\hspace{0.6cm}
L = -\frac{2BQ}{r}+\frac{1}{2}B^3Qr,
\end{equation}
\begin{equation}
\mathcal{Z} = 1+\frac{Q^2}{r^2}-\frac{2}{r},
\hspace{1.5cm}
P = B^2r+\frac{B^4r^3}{4},
\end{equation}
\begin{equation}
\begin{aligned}
S &= \frac{B^3Q}{2}+\frac{2BQ}{r^2},
\hspace{1.5cm}
K &= -\frac{2Q^2}{r^3}+\frac{2}{r^2},
\end{aligned}
\end{equation}
\begin{equation}
F = -\frac{r^2P}{\mathcal{A}^2}+\frac{2r}{\mathcal{A}},
\end{equation}

\begin{equation}
W = -\frac{r^2LP}{\mathcal{A}^2}
+\frac{Sr^2}{\mathcal{A}}
+\frac{2rL}{\mathcal{A}},
\end{equation}
\begin{equation}
J = -\frac{r^2L^2P}{\mathcal{A}^2}
+\frac{2Sr^2L}{\mathcal{A}}
+\frac{2rL^2}{\mathcal{A}},
\end{equation}
\begin{equation}
\mathcal{D} = W^2+F\left(\mathcal{Z}P+K\mathcal{A}-J\right),
\end{equation}
\begin{equation}
\mathcal{T} = -\frac{L}{\sqrt{\mathcal{A}\mathcal{Z}}}\sqrt{\frac{r^2}{\mathcal{A}}},
\end{equation}
\begin{equation}
\mathcal{U} = \frac{1}{\sqrt{\mathcal{A}\mathcal{Z}}\,F}
\sqrt{\frac{r^2}{\mathcal{A}}}
\left(W+\sqrt{\mathcal{D}}\right)\, .
\end{equation}

\bibliographystyle{apsrev4-1}  
\bibliography{gravreferences,Ref_BS1,Ref_BS2}

\end{document}